\documentclass[preprint2]{aastex}
\usepackage{aastexug}
\newcommand{\etal}{et~al.\ }
\newcommand{\HI}{\hbox{{\rm H}\kern 0.1em{\sc i}}}
\newcommand{\HII}{\hbox{{\rm H}\kern 0.1em{\sc ii}}}
\newcommand{\OII}{\hbox{[{\rm O}\kern 0.1em{\sc ii}]}}
\newcommand{\NII}{\hbox{[{\rm N}\kern 0.1em{\sc ii}]}}
\newcommand{\SII}{\hbox{[{\rm S}\kern 0.1em{\sc ii}]}}
\newcommand{\OIII}{\hbox{[{\rm O}\kern 0.1em{\sc iii}]}}
\newcommand{\kms}{\hbox{km~s$^{-1}$}}

\def\msun{\hbox{$\hbox{M}_{\odot}$}}

\newcommand{\hst}{{\emph{HST}}}
\newcommand{\simgt}{\lower 2pt \hbox{$\, \buildrel {\scriptstyle >}\over {\scriptstyle\sim}\,$}}
\newcommand{\simlt}{\lower 2pt \hbox{$\, \buildrel {\scriptstyle <}\over {\scriptstyle\sim}\,$}}

\tighten
\begin{document}
 
\shorttitle{\hst\ IMAGES OF STEPHAN'S QUINTET}
\shortauthors{GALLAGHER ET~AL.}

%%%%%%%%%%%%%%%%%%%%%%%%%%%%%%%%%%%%%%%%%%%%%%%%%%%%%%%%%%%%%%%%%%%%%%%%%%%%%%%%%%

\title{{\emph{Hubble Space Telescope}} Images of Stephan's Quintet: Star Cluster Formation in
a Compact Group Environment\altaffilmark{1,2}}

\author{Sarah~C.~Gallagher, Jane~C.~Charlton\altaffilmark{3}, Sally~D.~Hunsberger}
\affil{Department of Astronomy and Astrophysics \\ 
       The Pennsylvania State University \\
       University Park PA 16802 \\ 
       {\it gallsc, charlton, sdh@astro.psu.edu}}

\author{Dennis Zaritsky}
\affil{Steward Observatory \\
       University of Arizona \\
       Tucson, AZ 85721 \\
      {\it dennis@as.arizona.edu}}
\and
\author{Bradley C. Whitmore}
\affil{Space Telescope Science Institute \\
       3700 San Martin Drive\\
       Baltimore, MD 21218 \\
      {\it whitmore@stsci.edu}}

\altaffiltext{1}{Based on observations obtained with the
NASA/ESA {\it Hubble Space Telescope}, which is operated by the STScI
for the Association of Universities for Research in Astronomy, Inc.,
under NASA contract NAS5--26555.}
\altaffiltext{2}{Based on observations obtained with the Hobby--Eberly
Telescope, which is a joint project of the University of Texas at Austin,
the Pennsylvania State University, Stanford University, 
Ludwig--Maximillians--Universit\"at M\"unchen, and
Georg--August--Universit\"at G\"ottingen.}
\altaffiltext{3}{Center for Gravitational Physics and Geometry}

\begin{abstract}
Analysis of {\it Hubble Space Telescope}/Wide Field Planetary Camera 2
images of Stephan's Quintet, Hickson Compact Group 92, yielded $115$ 
candidate star clusters (with $V-I < 1.5$).  Unlike in merger remants, 
the cluster candidates in Stephan's Quintet are not clustered in the inner regions of 
the galaxies; they are spread over the debris and surrounding area. 
Specifically, these sources are located in the long sweeping tail and
spiral arms of NGC~7319, in the tidal debris of NGC~7318B/A, and in the
intragroup starburst region north of these galaxies.  
Analysis of the colors of the clusters indicate several distinct epochs of star 
formation that appear to trace the complex history of dynamical interactions 
in this compact group. 
\end{abstract}

\keywords{galaxies: interactions --- galaxies: individual (NGC~7318A, NGC~7318B, NGC~7319)
--- galaxies: star clusters --- intergalactic medium}

%%%%%%%%%%%%%%%%%%%%%%%%%%%%%%%%%%%%%%%%%%%%%%%%%%%%%%%%%%%%%%%%%%%%%%%%%%%%%%%%%%

\section{Introduction}
\label{sec:intro}

The interactions of galaxies trigger bursts of star formation, from
the compact star clusters formed around the central regions of mergers
\citep{Holtz92,whitmore93,schweizer96,miller97,whitmore99,zepf99} 
to the dwarf galaxies along and at the ends of their
tidal tails \citep{mirabel,sally96}.
Theoretical considerations motivated the prediction that there would
be multiple generations of clusters based upon the sequence of merger
events \citep{ashman92,zepf93}, and
specific conclusions can be drawn about the interaction history of a
galaxy pair from detailed studies of its populations of compact
star clusters.  A compact group of galaxies thus offers the possibility of an even
richer laboratory for such studies.  In particular, the
high densities coupled with low velocity dispersions ($\sigma\approx200$--300~\kms)
of compact groups make them active sites of strong galaxy interactions, probably
similar to events in the early universe.

Arguably the most famous and well-studied merger is ``the Antennae'' (NGC~4038/39)
which has regions in which very young ($\simlt5$~Myr) and young ($5$--$10$~Myr) 
clusters are found.  In the same system, there are also intermediate-aged populations
with ages of $\sim100$~Myr and $\sim500$~Myr, the latter apparently
related to the same interaction that generated the tails \citep{whitmore99}.
Finally, there is an old population of clusters presumably formed with
the galaxies themselves.  In contrast, NGC~3921 and NGC~7252 do not have 
populations of very young clusters; their intermediate-aged populations
are consistent with forming along with their tidal tails \citep{schweizer96,miller97}.  
Some mergers have ongoing star cluster formation
continuing well after the major event, but in others the rate of cluster formation 
appears to have decreased.

The young compact clusters that are detected in mergers range in luminosity
from $-14 \lesssim M_V < -9$ \citep[e.g.,][]{schweizer96}.  
Fainter objects are found in deep images of
NGC~4038/9, but distinguishing these from individual
supergiant stars is difficult \citep{whitmore99}. 
The luminosity function of the young cluster population roughly
follows a power law in contrast to that of old globular cluster systems,
which is a lognormal luminosity function with a peak at
intermediate luminosities \citep[e.g.,][]{harris91}.

Young star clusters are also found in the tidal debris of some, but
not all mergers.  Knierman \etal (2001) found populations of young
($\sim 5$~Myr) and intermediate-aged ($30$--$300$~Myr) clusters in the
debris of NGC~3256.  However, they did not find significant numbers
of compact clusters in the debris of the three other mergers studied
(NGC~7252, NGC~3921, and NGC~4038/39).  They hypothesized that star
clusters may form in environments where larger condensations, such as
tidal dwarf galaxies, are not able to form, but emphasize that larger
samples are needed to confirm this hypothesis.  
The overall star formation efficiency is likely to be influential
in determining the compactness and stability of a cluster.  High
pressure caused by some external trigger may be needed to form a
massive globular cluster \citep[e.g.,][]{elmegreen2000}.
It is important to collect more empirical data to better
understand the initial conditions required for formation of clusters
of various sorts.

Hickson compact groups (HCGs) provide a fertile, if complicated, environment
in which to study the relationships between interaction and merger
events and the ensuing formation of star clusters.
The HCGs were selected
as groups of $\ge4$ galaxies on the basis of compactness and isolation
from other bright galaxies \citep{hickson82}.  Their high galaxy densities 
are comparable to those in the centers of galaxy clusters. However,
given the median velocity dispersion of $\sim200$~{\kms}, a larger fraction of 
interactions will lead to mergers than in the galaxy cluster environment.
Enhanced interaction rates are supported by the deficiency of CO in
HCG spirals (presumably due to stripping) and by enhanced FIR emission
due to nuclear starbursts \citep{verdes98}.  Extended X-ray halos
\citep{saracco95,ponman96} also provide evidence for enhanced
rates of interactions, as do the number of tidal features presently
found in the HCGs \citep{sally96}. There should be a record preserved 
in the stellar populations of structures formed in various events. 

In Hickson Compact Group 31 (HCG~31), a detailed study of the
star formation history has been possible using {\it Hubble Space Telescope} (\hst)
$B$, $V$, $R$, and $I-$band images \citep{johnson99} and WIYN H$\alpha$ images.
Although the major interaction occurred $\sim400$~Myr ago, there is evidence
for recent ($1$--$10$~Myr) star formation throughout the group \citep{johnson00}.  
Johnson and Conti (2000)\nocite{johnson00} speculate that one of the smaller
galaxies may itself have been formed only $\sim4$~Myr ago in tidal debris.
They also find evidence in other regions of old stellar populations that have
presumably been tidally removed from the galaxies.  However, there are
relatively few star clusters older than $10$~Myr, and although
clusters fade, it is unlikely that the brightest
of them would have faded below the detection threshold of that study.

Our similar study, presented here, focuses on HCG~92, Stephan's Quintet (hereafter, SQ).
This group was discovered more than $120$ years ago \citep{stephan}
and was later included in the HCG catalog \citep{hickson82}.
An $R-$band image of the group, obtained with the 1.5~m Palomar telescope,
is presented in Figure~\ref{fig:palomar-fig}, with the galaxies and key areas
labeled. NGC~7320 has been shown to be a foreground galaxy at
$\sim 800$~{\kms} and therefore only a chance superposition with the 
physical group \citep{claudia94}.
The three galaxies, NGC~7317, NGC~7318A, and NGC~7319 are all within
$50$~{\kms} of $6600$~{\kms}.  However, the galaxy NGC~7318B,
at first glance apparently closely associated with NGC~7318A, is
in fact blueshifted by $900$~{\kms} relative to the others \citep{claudia94}.
The redshift of the group is $z=0.0215$ which places it at $D\sim85h_{75}^{-1}$~Mpc.
At this distance, a pixel on the Wide Field Camera covers 
$\sim39h_{75}^{-1}$ pc, and so a compact star cluster is not
resolved.

Three of the galaxies in SQ (NGC~7318A, NGC~7318B, and NGC~7319) show 
signs of morphological irregularities, however, NGC~7317 appears
undisturbed.  SQ also has the richest known system of tidal dwarf galaxy candidates,
with thirteen in the prominent tail to the west of NGC~7319, several in the eastern
spiral arm of NGC~7318B, and several in the northern starburst region
\citep{sally96}.  Recent star formation is evident in several locations in
ground-based $B-V$ \citep*{schombert90}, H$\alpha$ \citep{vilchez98}, and 
far-IR images \citep{xu99}.

Moles, Sulentic, \& M\'arquez (1997)\nocite{moles97} construct a history of SQ 
in which the fainter galaxy to the northeast, NGC~7320C, passed through the
group a few hundred million years ago.  The interaction with that galaxy
stripped NGC~7319 of much of its gas \citep{shostak}, some of which remained
as a reservoir for the northern starburst region.  This event could also be 
responsible for the large tidal tail extending to the east from NGC~7319.
According to this scenario, bursts of star formation would have occurred during 
the period of time surrounding this interaction.
However, Xu \& Tuffs (1999)\nocite{xu99} have used ground-based H$\alpha$
and near-IR images as well as {\it Infrared Space Observatory} ({\it ISO})
mid-IR images to determine that at least
some of the star formation in the northern starburst region must have occurred
much more recently, perhaps $10$--$20$ million years ago.  This young burst would have
been triggered by the rapid passage of the galaxy NGC~7318B through
that region.

This article presents $B$, $V$, and $I-$band images of Stephan's Quintet, 
obtained with the Wide Field Planetary Camera 2 
(WFPC2) aboard \hst.
The star formation histories in the various regions of this compact
group are studied through our analysis of the colors of populations of 
compact star clusters.
The goal is to determine which of the evolutionary events triggered formation of these
clusters globally, and/or in local regions throughout the group.
In \S~\ref{sec:obs} the observational details are given, the
reduction, detection and photometry techniques are outlined, and the images
are presented.
Color-magnitude and color-color diagrams are presented in
\S~\ref{sec:results}, the different star cluster populations
are identified in the images, and the ages of these populations
are discussed.
Long-slit spectra of some prominent star-forming clumps, obtained with
the Marcario Low Resolution Spectrograph on the Hobby-Eberly Telescope
(HET), are presented in \S~4.
A summary of conclusions and general discussion comparing the cluster
population in Stephan's Quintet to those in other environments
follow in \S~\ref{sec:conclude}.

%-----------------------------------------------------------------------------------------
\section{Observations and Data Analysis}
\label{sec:obs}

Stephan's Quintet was observed with the WFPC2 aboard \hst\ in 
two pointings.  Field 1, the first set of images taken on 1998 December 30, encompassed
the southern spiral arm of NGC~7319, NGC~7318B/A, and the northern starburst region.  
The second, Field 2, taken on 1999 June 17, covered NGC~7319 and its extended 
tidal tail as well as a ``sky'' region containing no obvious tidal features.  
The images in the PC overlapped with WFC images, and we consider only the latter 
in the following analysis as they were more sensitive for point source detection.
On both occasions, the images were once dithered  
and taken through three wide-band filters: F450W ($B$), F569W ($V$), and F814W ($I$). 
Transformations from these \hst\ filters into the standard
Johnson $B$ and $V$ and Cousins $I$ filters can be found in \citet{Holtz95}.
This choice of wide-band filters was motivated by the goal of reducing contamination 
to the colors from common bright nebular emission lines such as \OIII\ and H$\alpha$.
At the observed  wavelength of \OIII\ emission from Stephan's Quintet, 
$\lambda_{\rm obs}\approx5114$~\AA, 
the transmissivity of F569W is $\approx73\%$ as opposed to $\approx83\%$ for the usual visible
F555W filter. 
The transmissivity of F450W is $\approx78\%$ whereas the more standard blue F439W filter has 
essentially no throughput at this wavelength.  The close match in transmissivity of 
F569W and F450W thus serves to reduce the \OIII\ contribution to $B-V$ color.  
None of our filters has notable sensitivity to H$\alpha$ emission, 
$\lambda_{\rm obs}\approx6704$~\AA.

Dithering entails offsetting the image position
by $\approx2.5$~pix in the WF camera in both the $x$ and $y$ directions to increase 
the effective resolution of the combined image by better sampling the point spread function
(PSF).  In this case, we obtained two images at each position in each field and filter.
The exposure times in each field were $4\times 1700$~s, $4\times800$~s, and 
$4\times500$~s for $B$, $V$ and $I$, respectively.  
The gain in each case was 7 e$^-$/ADU.  The data were first processed through the 
standard \hst\ pipeline.  Subsequently, they were cleaned of cosmic
rays using the IRAF{\footnote{IRAF is distributed by the National Optical 
Astronomy Observatories, which is operated by the Assocation of Universities 
for Research in Astronomy, Inc. under
cooperative agreement with the National Science Foundation.}} 
task GCOMBINE, which averaged the two images taken in each position,
followed by the task COSMICRAYS to remove hot pixels.
Figure~{\ref{fig:sq-all-hst} shows the cleaned $V-$band image of both fields 
combined with the regions of interest labeled.

\subsection{Object Detection and Classification}
The initial detection of point sources in each image
was undertaken using the DAOFIND routine in DAOPHOT \citep{stetson}
with a very low detection threshold of $\sigma=2$ for the peak amplitude of the
Gaussian.
The background over the entire
field of SQ varies considerably, and this strategy was chosen following Kundu \etal (1999)
\nocite{kundu99} to aid in detection
of sources in regions with high background.
This technique produced thousands of sources per chip, and 
we then performed aperture photometry using the task PHOT in the package APPHOT
on all candidate sources.  The aperture 
radius was 2 pix ($0.^{\prime\prime}19$), 
and the background was the median value in an annulus around
the aperture with inner radius of 5 pix and width of 3 pix.
Those sources with a signal-to-noise ratio above a fixed level, $S/N>3.0$, that 
appeared in each filter at both dither positions were retained.  The noise for each source
was calculated from the Poisson noise of the target and the uncertainty in the background, 
combined in quadrature.

After this step, many spurious sources remain ``detected'' in particularly 
noisy regions such as diffraction spikes around bright stars and dust lanes in 
galaxies.  At the distance of SQ, $85h_{75}^{-1}$~Mpc, 1~WFC pixel corresponds
to $39h_{75}^{-1}$ pc which is larger than the expected effective radius of even 
large star clusters 
($R_{\rm eff}<15$~pc, e.g., Whitmore \etal 1999\nocite{whitmore99}).  
Therefore, we expect all star clusters to be point sources.
To eliminate irregular or extended structures, we followed
the example of Miller \etal (1997)\nocite{miller97}. 
We calculated $\Delta V_{0.5-3.0}$, the difference between the $V$ magnitudes 
from two photometric apertures: one with a radius of 0.5~pix and the other 
with a radius of 3.0~pix. Those sources
with $\Delta V_{0.5-3.0}>2.4$ can be classified as extended while those with
$\Delta V_{0.5-3.0}<0.5$ are probably residual cosmic rays or hot pixels missed in the 
cleaning process.  In either case, such sources are considered 
unlikely star cluster candidates. 

In addition to this test, the radial profile of each source was fit with
a cubic spline of order 5 using the APPHOT task RADPROF. 
From an examination of the distribution of FWHM of these profile fits, 
point-like candidates have $0.5\le{\rm FWHM}\le2.5$~pix.  The upper limit
is the same as found by Miller \etal (1997) using Gaussian models, but we
have extended their lower limit from 1~pix to 0.5~pix 
based on visual inspection of the candidates.

For the final determination of star cluster candidates, 
we compiled $\Delta V_{0.5-3.0}$ and the radial profile
FWHM of each source in the $V-$band images taken at each dither position. 
To be kept as a candidate, a source had to pass three out of the
four (two for each dither position) point-source classification tests.
This resulted in a total of 458 star cluster candidates, 276 in 
Field~1 and 182 in Field~2.  Of those objects satisfying the morphological
criteria, 189 were found in all three filters.  For our final candidate list,
only those sources with color errors, $(B-V)_{\rm err}$ and $(V-I)_{\rm err}$, 
less than 0.2 mag were retained.  The final list contains 139 point sources
whose positions are indicated in Figure~{\ref{fig:sq-all-hst}}.  
Of these, 24 have $V-I>1.5$ which places them outside of the range of likely
star clusters; these sources are probably foreground stars in our Galaxy \citep{schweizer96}.
Though young star clusters embedded in dust can have colors as red as $V-I\approx4$
\citep[e.g.,][]{whitmore99}, few if any of our sources with $V-I>1.5$
appear to be found in the dust lanes, hence this is not likely to be a significant problem.

This sample will clearly contain some foreground stars and background galaxies, 
but the spatial coincidence of most of the sources with the galaxy bulges, disks, and 
tidal features suggests that many candidates are legitimate star clusters. 

\subsection{Completeness Limits}
\label{sec:completeness}
To estimate the completeness of our star cluster candidate sample, \citet{whitmore99} was
used as a methodological guide.  We focused on the region containing NGC~7318B/A where the
background level is structured and variable, and possible selection effects could influence 
our findings.  The $V-$band images of NGC~7318B/A were populated with artificial point 
sources using the ADDSTAR task within DAOPHOT.  Our detection algorithm was then 
followed to determine which fraction of the added point sources were recovered as a 
function of background level.  The background level contours of a cropped region of the field 
and the completeness fraction versus apparent $V$ magnitude are shown in 
Figure~\ref{fig:completeness}.  The majority
of the area of Stephan's Quintet has background values $<7$~DN.  Even up to background values
$\lesssim50$~DN (encompassing most of the area of the galaxy bulges), 
our sample is $\approx50\%$ complete to an apparent magnitude of 25.1, equivalent to $M_V=-9.6$.

\subsection{Final Photometry}
\label{sec:finalphot}
After the final selection detailed above, we performed circular aperture photometry on 
all of the detected objects using a 
2~pix radius aperture.  Though a larger aperture would collect a greater fraction
of the total light, a smaller aperture for faint sources preserves a higher 
signal-to-noise ratio.
The sky level is taken to be the median of an annulus with a width of 3~pix 
and a 5~pix inner radius.  
For each point source, the photometry was performed independently at each dither position and
then averaged.  These data were also corrected for charge transfer efficiency losses 
according to the prescription in Whitmore, Hayer, \& Casertano (1999)\nocite{whc99}, 
but due to the long exposures (and therefore high background) these corrections were
not significant.

The zeropoints in the Vega magnitude system for these filters have been taken from
the \hst\ Data Handbook.  These zeropoints (for infinite aperture), have been
calibrated to be 0.1 mag less than the zeropoint at $0.^{\prime\prime}5$.  Therefore,
the aperture correction from 2 pix to $0.^{\prime\prime}5$ (5.15 pix) must be determined.  
Following Kundu \etal (1999), aperture photometry for the brightest, isolated point sources in 
each field and chip was performed with several aperture radii from 0.5 to 6 pix.  The number 
of appropriate bright sources available for each filter ranged from 9 to 19.  The 
difference in magnitude between an aperture with a 2.0 and 5.15 pix radius was the 
mean of that measured for all the point sources used.
There were not enough point sources per WFC to
determine the aperture correction independently; in any case, the differences across an 
individual chip are generally greater than between chips.  
The mean aperture corrections are
$-0.194\pm0.042$ (F450W), $-0.200\pm0.030$ (F569W), and $-0.236\pm0.023$ (F814W) for
Field 1, and $-0.286\pm0.051$ (F450W), $-0.278\pm0.039$ (F569W), and $-0.334\pm0.036$ (F814W)
for Field 2.  The errors are the standard deviations in the mean aperture corrections.
Because the two pointings were
taken approximately 6 months apart, the aperture corrections
for each field are different as a result of differing focus.  
This change in focus arises from the ``breathing'' of the 
telescope; the focus is reset approximately every six months.

From Burstein and Heiles (1984), the foreground Galactic extinction towards SQ is
$A_B=0.33$ which gives $E(B-V)=0.083$.  Using the synthetic photometry package
SYNPHOT in IRAF, synthesized Bruzual \& Charlot instantaneous-burst star cluster 
spectra (assuming a Salpeter initial mass function and solar metallicity) 
were reddened using a Seaton extinction law \citep{seaton79}.
These reddened spectra were then convolved with the \hst\ filter response to determine
the reddening correction for each filter: $A_{\rm F450W}=0.325$, $A_{\rm F569W}=0.254$, and
$A_{\rm F814W}=0.161$.  The reddening corrections were consistent within 0.01 mag for
star cluster spectra with ages from 1~Myr to 10~Gyr.  All of the photometry in this paper
has been corrected for Galactic reddening.

\section{Results}
\label{sec:results}

Figures~{\ref{fig:cc-fig}} and {\ref{fig:cmd-fig}} contain the $B-V$ versus $V-I$ color-color
and the $V-I$ versus $V$ color-magnitude diagrams 
of all of the point sources meeting the detection criteria detailed above.
In addition to Galactic reddening, internal dust extinction is also likely to affect the
colors of these sources.  However, this internal reddening is likely to be highly variable
considering the diversity of environments in which we find star cluster candidates. 
For reference, the color-color diagram has a reddening vector 
included which represents 1.0 mag of extinction in the Johnson $V$ filter; we use 
the Johnson, rather than \hst, $V$ filter for consistency with previous work.
Histograms of the $V-I$ and $B-V$ colors are displayed in Figure~\ref{fig:colorhis}.
Sources redder than $V-I = 1.5$ are likely to be foreground stars, though we may be
misclassifying some dust-enshrouded clusters.  Excluding
those sources, the luminosity function of the star cluster candidates is
given in Figure~{\ref{fig:lf-fig}.  A maximum likelihood power-law fit 
to the luminosity function yields $N(L) \propto L^{-2.0\pm0.1}$ 
to a cutoff magnitude of $-10.2$.
The power-law slope is not very sensitive to the exact 
value of the cutoff between $-10.6<M_V<-10.0$.   
In any case, because the clusters are unlikely to be a
population with a single origin, it is unclear how this model fit should be interpreted.
There were not enough objects within different populations to fit them individually.

The colors of clusters help us to sort out ambiguities between the 
masses/luminosities of clusters and their ages, though intrinsic reddening
can always make clusters appear older than they actually are. 
In general, we expect clusters to become redder and to fade as they become older.  
Figure~\ref{fig:histograms} suggests such a trend, with a larger fraction
of fainter clusters as the colors become redder, but there is a large
amount of apparent ``contamination'' due to apparently bright, red
objects.

Figure~\ref{fig:sq-all-hst} shows cluster candidates spread over the
area of the group.  Because of the complex group history and multiple
interaction events, we now consider each of the marked regions separately.

\subsection{Tidal Tail}
\label{sec:reg-tidal}

Within the tidal tail region in Figure~\ref{fig:tail-figs}, $23$ point sources
 were identified.  Of these, six have red colors ($V-I>1.5$) more
typical of stars than star clusters; some of these also have $V$ magnitudes
that make them unreasonably luminous for red clusters at the
distance of Stephan's Quintet.  They are indicated in
Table~\ref{tab:tail-table} with an ``S'' in the last column and are excluded
from the following discussion.
The physical association of all of the non-stellar sources, except perhaps objects 1 and 8,
with the optical tail supports their identification as star clusters associated
with the group.

In the color-color diagram for this region (Figure~{\ref{fig:tail-figs}}b),
there is evident clustering along the evolutionary track in the age range
$\sim10-500$ Myr.  
Unfortunately, with the $B-V$ versus $V-I$ color-color diagram it is not
possible to accurately determine an age within this range.
The cluster candidates that lie above the track,
at $B-V \sim 0.4$ and $V-I \sim 1.1$, have positions consistent with
ages of a few tens of Myr and $\sim0.75$ magnitudes of extinction; however
the degeneracy between age and reddening makes such a claim uncertain.
If significant reddening affects several of the sources consistent
with unreddened ages of $\sim10$~Myr, it is possible that there could
be very young ($<5$~Myr) star clusters in the tail, but we have no
direct evidence for such a population.

It is interesting to compare the ages of clusters in the tail to the age
of the tidal tail itself.  Through such a comparison we can consider
whether the clusters were drawn out from the parent galaxy, or whether they formed
in the tail itself.
The age of the NGC~7319 tidal tail has been estimated by Moles \etal
(1997\nocite{moles97}) by assuming that the tail formed with the
most recent encounter of NGC~7320C with NGC~7319.  Briefly,
the velocity of NGC~7320C in the transverse direction was assumed
equal to that in the radial direction.  At this velocity,
it would travel to its $140h_{75}^{-1}$~kpc separation on the sky in
$2.0 \times 10^8 h_{75}^{-1}$~yr.  Another rough estimate for the age
of a tidal tail is given by the ratio of the length of the tail ($\sim30h_{75}^{-1}$~kpc)
to the rotational velocity of the galaxy ($\sim200$~\kms) which gives
$1.5 \times 10^8$~yr, consistent with the estimated time of last
encounter with NGC~7320C.  Clearly, these are rough estimates, but
we conclude that the ages of most of the star clusters in the tail of NGC~7319
are consistent with their formation within the tail itself.  It is also 
possible that several of them are actually younger.

The majority of the clusters are not considerably older than the tail,
as they would be if they formed with NGC~7319 itself.
However, two objects, 1 and 21, have colors consistent with ages of a few Gyr.
Number 21 is close to the base of the tail, and could easily be an outlier
from the parent galaxy globular cluster population.  Number 1 would have had to be
pulled out with the tail, but it does not lie directly in the tail.
Object 3 appears significantly to the left of the tail on the color-color
diagram.  This may be a compact background galaxy or perhaps
a young cluster with a nebular contribution to its colors
(see e.g., Johnson \etal 1999\nocite{johnson99}).

Within our photometric errors, there is no obvious relationship between
the implied ages (colors) of the star clusters and their locations along the
length of the tail; star cluster candidates at the tip (e.g., objects 2
and 4) and those nearest to NGC~7319 (e.g., object 22) occupy the same space in
the color-color diagram. 

Object 12 is notably luminous with $M_V=-13.2$.
This $M_V$ is consistent with the luminous tail of the distribution of 
star clusters found in the inner regions of ongoing young mergers such
as the Antennae \citep{whitmore99} and NGC~7252 \citep{miller97}, and it
is also quite blue, with $B-V = 0.14$.  However, this is the first reported 
detection of such a luminous cluster
in a tidal tail.  Given its relative brightness ($V=21.53$), this object
is an excellent target for future spectroscopy to confirm its identification
and constrain its age and metallicity. 

\subsection{Sky Region}
\label{sec:reg-sky}

An expanded view of the sky region, with its $13$ point sources identified,
is presented in Figure~\ref{fig:sky-figs}, along with the locations of the
sources in a color-color diagram.  Three of them have red colors consistent
with stars.  The colors and magnitudes of the individual sources are listed
in Table~\ref{tab:sky-table}.

The sky region could be considered as the only ``blank'' region in the two
WFPC2 visits to SQ, and therefore could in principle be used to statistically
remove background contamination from our star cluster counts.
However, the concentration of the three bluest sources (objects 30, 31, and 33)
in the same region of the images suggests that perhaps there are some young
star clusters here as well.  The implied ages are consistent ($\sim10$--$500$~Myr) with
the NGC~7319 tail population.
Star clusters at the upper end of this age range could have formed
in the outer regions of NGC~7319 and then been stripped out into the
intergalactic medium.  We estimate a minimum timescale for this
of $\sim100h_{75}^{-1}$ Myr for such a cluster to travel 
$20h_{75}^{-1}$~kpc at a typical galactic  
escape/rotation velocity of $\sim200$~\kms.

Alternatively, the formation of clusters in this intergalactic region is not implausible
as it contains a significant amount of {\HI} \citep{shostak} as well as diffuse, blue
optical emission \citep{schombert90}. The existing intergalactic \HI\ could be 
compressed by an encounter so that its density is increased, thus enhancing the star formation
efficiency \citep{elmegreen2000}.

\subsection{NGC~7319 and Vicinity}
\label{sec:reg-7319}

NGC~7319, classified as a barred spiral \citep[Sbc;][]{hickson92} is the
member of SQ with the most prominent evidence of tidal disturbance.
There are $33$ point sources, $8$ of them likely to be stellar with $V-I > 1.5$,
in this region.  The color-color diagram for these sources is presented
and the individual sources are identified in the image in
Figure~\ref{fig:7319-panel}.

Most of the bluest cluster candidates in the region, with ages $8$--$100$~Myr
(triangles and squares in Figure~\ref{fig:7319-cc}), are located along the
double spiral arm. It is interesting that these clusters may have formed
well after any known interaction of NGC~7319 with NGC~7320C.  As discussed in
\S~\ref{sec:reg-tidal}, the last encounter with NGC~7320C is likely
to have occurred $\sim200 h_{75}^{-1}$~Myr ago.  
Perhaps more recent cannibalism of a satellite is responsible for this star formation event.
Alternatively, late formation of star clusters could have been triggered by a delayed
return of tidally ejected material over a timescale of a few Gyr \citep{hibbard95}.

Intermediate-aged and older clusters, including several
consistent with ages of $\sim10$~Gyr, are distributed more uniformly throughout
the region.  There are no luminous ($M_V < -10$) clusters in the inner regions
of NGC~7319.  This is in contrast to the classical Toomre sequence mergers,
such as NGC~7252 \citep{miller97} and NGC~3921 \citep{schweizer96}, in which
young, blue clusters are abundant in the central regions.  
Given our completeness limits, such a population should have been 
detectable for distances $\gtrsim2h_{75}$~kpc from the galaxy nucleus unless 
embedded within dust clouds.
In addition, some point sources are detected 
in the bulge (e.g., objects 54, 58, and 59) with absolute magnitudes down to $-10.7$.  

Several of the older cluster candidates (with $B-V\sim0.85$ and $V-I\sim1.15$)
are very luminous ($M_V<-13$) if at the distance of Stephan's Quintet.
A cluster with such red color and a solar metallicity would have $M_V<-13$
only if it was more massive than $\sim4\times10^7$~{\msun} \citep{schweizer96}.
Even with sub-solar metallicity the masses would be $\gtrsim10^7$~{\msun}, 
considerably higher than for clusters seen in other
environments.  Perhaps these are foreground stars or the concentrated nuclei of dwarf elliptical
galaxies rather than isolated star clusters.  If so, they significantly
contaminate the overall luminosity function presented in Figure~\ref{fig:lf-fig}.
As the reddening vector is almost parallel to the evolutionary
tracks, extinction can cause clusters to appear older than their actual
ages.  However, these bright star cluster candidates are located at the 
outskirts of NGC~7319 where substantial extinction is not expected.

\subsection{NGC~7318 B/A and the Northern Starburst Region}

The region surrounding NGC~7318 B/A is the most fertile site for star cluster candidates.
Seventy point sources, more than half of the total number in the entire
field, were found in this region (shown in Figure~\ref{fig:7318-panel}).
The sources, only $8$ of which we classify as stellar, are listed in
Table~\ref{tab:7318-table}.  Many are associated with bright, blue structures
such as the eastern spiral arm of NGC~7318B and the northern tidal debris.  

This region is also the only one of the four to contain very young cluster candidates
with ages $<5$~Myr, see Figure~{\ref{fig:7318-cc}}.  
These clusters are not localized in a single burst region.
They are spread over this area, both north and south of NGC~7318 B/A, and extending
over $>40h_{75}^{-1}$~kpc.  Objects 90 and 93 are both luminous
($M_V = -13.01$ and $-13.43$) and extremely blue, consistent with clusters
of mass $\approx3\times10^5$~{\msun}.
Another striking feature of the color-color diagram for this region is the tight
knot of $11$ star cluster candidates coincident with $\approx7$~Myr on the 
evolutionary track.  This population (object 69 in Table~\ref{tab:7319-table} and 
indicated with asterisks in Table~\ref{tab:7318-table}) is physically associated 
with both NGC~7318B and the
northern starburst region.  Such coeval star formation on physical scales up to 
$\approx40$~kpc is remarkable.

The northern starburst region hosts a luminous far-infrared $ISO$ source with
bright H$\alpha$ emission \citep{xu99} which is coincident with a dwarf galaxy.  
We find several young cluster candidates 
within the immediate environs of this source (see Figure~{\ref{fig:dwarf-fig}}).  
Based only on the color-color age
determination, these point sources appear to encompass a range of ages.  However,
extinction is likely to be significant and patchy in such an active region,
making these determinations unreliable.  Therefore, these ages should be considered 
upper limits as reddening can only make a cluster appear older.  
The northern starburst region is
of particular interest because active star formation has apparently occurred more
recently than the past major encounter.  Assuming that NGC~7318B is traveling
at $900$~\kms\ it has been at least $20$~Myr since it passed through the
northern starburst region.  From the ratio of the stellar mass to the near-IR
luminosity Xu {\etal} (1999) estimated a starburst age of $10$--$20$~Myr.
We find several star clusters younger than this, and therefore they must
have formed after the initial burst and the actual passage of NGC~7318B through
the region.

\section{Spectra of the Tidal Clumps of NGC 7318B}

A long-slit spectrum of a region of tidal debris south of NGC 7318B was obtained on
1999 November 11 with the Marcario Low-Resolution Spectrograph \citep{lrs1,lrs2}
on the Hobby-Eberly Telescope \citep{het}.  This particular region has at least
five separate areas that can be distinguished along the long slit.
A $600$ line mm$^{-1}$ grism was used with the $2$\arcsec\ slit to give $R\sim600$
over a 4-pixel resolution element.  The spectrum covered the wavelength range
$4000$--$7000$~{\AA}.  At the wavelength of H$\alpha$ at $z=0.0215$, a pixel
corresponds to $2.05$~{\AA}, or $\sim 92$~{\kms}.  Wavelength calibration was
provided by Ne, Cd, and Ar comparison lamps.

The position of the slit is illustrated on a portion of the \hst/WFPC2 image in
Figure~\ref{fig:het} and the image of the full spectrum is shown.  A strong
continuum is apparent for the brightest blue clump, and a weak one for the
area closest to the top of the image, which contains two smaller blue knots.
In each of the emission lines, five separate debris regions can be distinguished.
Approximate velocities are noted on the ``close--up'' view of the wavelength
region covering H$\alpha$, {\NII}, and {\SII}.  We chose to display the
image of the spectrum rather than an extracted spectrum in order to illustrate
the complex and connected velocity structure in this region.

Comparing the velocities of the various regions along the debris to the 
velocities of the group galaxies may elucidate the origin of the gas in question.
From the top of Figure~\ref{fig:het} to the bottom:
The blue knots above the central clump of debris are $\sim150$--$300$~{\kms}
blueward from the strong continuum source (at $~5700$~{\kms}).  The latter
is consistent with the velocity of NGC~7318B, which is blueshifted by
$900$~{\kms} from the rest of the group.  The strong continuum source also has
strong emission lines and both are thought to come from the tight blue knot
seen in the \hst\ image.  However, interpretation is confused by the superposition
of this blue knot on a much redder galaxy.  This redder galaxy could be a dwarf member of the
group but its clear spiral structure suggests that it may just be a superposition
of the debris on a background galaxy.  The extracted spectrum from this region
does not show any spectral features that would indicate the redshift of this
galaxy.  The region of the spectrum below the strong continuum source shows
redshifted emission lines ranging from $350$--$800$~{\kms} redward of
$5700$~{\kms}.  There is emission close to the velocity ($6600$~{\kms})
of the three galaxies NGC~7317, NGC~7318A, and NGC~7319.

We conclude that there is emission from this debris at the velocities of both
of the galaxies NGC~7318A and NGC~7318B, as well as in between these two
velocities.  This seems not to be a single tidal feature, at least not one
destined to be long-lived.  Some of the material is leading NGC~7318B,
and some is trailing, with some material ``left behind'' at a velocity close to
that of NGC~7318A and NGC~7319.  There is also {\HI} at several velocities 
$5700$, $6000$, and $6600$~{\kms}, but only that at $5700$~{\kms}
is localized in this area south of NGC~7318B \citep{shostak}.
The northern starburst region shows components at $6000$ and $6600$~{\kms}
in {\HI} \citep{shostak}, in H$\alpha$ \citep{moles97}, and in CO \citep{smith},
indicating active star formation associated with gas at these two velocities.
Apparently, active star formation at different velocities is also present 
in the region of debris south of NGC~7318B.
This region contains a complex superposition of gas, not necessarily all associated with a
single physical structure.  It is consistent with a picture in which NGC~7318B
is flying at high speed through various ``layers'' of gas and triggering star
formation events, sometimes over large spatial scales.

\section{Conclusions and Discussion}
\label{sec:conclude}

The main conclusions of this work relate to the formation history of star
clusters in Stephan's Quintet (HCG~92).  Here we summarize the
conclusions and qualitatively compare the population of clusters (their ages 
and spatial location) to those in other environments.

\begin{enumerate}
\item{
There are $115$ point sources in SQ with $V-I < 1.5$, the vast
majority of which are likely to be star clusters.  These clusters have
$-14 < M_V < -9.5$, below which the completeness of
our study drops rapidly. Their colors are in the ranges $-0.3 < B-V < 1.0$
and $-0.2 < V-I < 1.5$.  
Color-color and color-magnitude diagrams
for the full sample of point sources, a histogram of colors, and 
the luminosity function of all star cluster candidates are presented.
}

\item{
We do not detect large numbers of star cluster candidates in the central
regions of the galaxies (NGC~7319, NGC~7318A, and NGC~7318B).  Instead, the
majority are in the tidal debris associated with the various galaxies, and
in the northern starburst region.  The colors of many of the candidate
clusters are blue, indicating that stars are apparently forming separate
from the disks and bulges of galaxies. 

Detection of bright, young clusters at large distances from galaxy
centers is in contrast to the young star cluster systems of isolated galaxy
pairs such as, for example, 
NGC~4038/39 \citep{whitmore99}, NGC~7252 \citep{miller97}, 
NGC~3256 \citep{zepf99}, and NGC~3921 \citep{schweizer96}.
Such pairs typically have hundreds of blue ($V-I < 0.8$) clusters
centered on the merger remnant.  Only one of the four pairs just listed,
NGC~3256, has an abundance of such clusters in its long, sweeping tails
\citep{knierman01}.}

\item{
Several distinct epochs of recent star formation have been identified in
various regions throughout SQ: (a) the northern starburst region contains
many of the very youngest clusters, some only $\sim2$--$3$~Myr
old.  Several such very young clusters are also found in the debris south of
NGC~7318B/A; (b) a group of more than a dozen clusters is found, all with colors
quite close to $B-V = 0.1$ and $V-I = 0.5$, which implies an age of $7$~Myr.
These objects, despite their similar ages, are distributed over
$>40h_{75}^{-1}$~kpc,
i.e., in both the northern starburst region and in the debris south of
NGC~7318B/A; (c) a group of clusters, both in the long tail of NGC~7319 and
surrounding the galaxy itself, with ages in the range $10$--$500$~Myr.
The $B-V$ versus $V-I$ color-color diagram does not allow accurate
measurements of the ages within this range; thus we cannot determine if
these clusters have a large range of ages or if they were formed over
one or more shorter intervals.  Most are consistent with having formed
in the tail itself, which is estimated to be about $150$~Myr old,
but some have blue colors that imply they may have formed more recently.
Regardless, these clusters were formed in the tail.  They are not an old
population of clusters drawn out from NGC~7319; (d) an old population ($3$--$12$~Gyr)
spread over the field, including some in the northern starburst region.  These
must have been stripped from one or more of the galaxies or formed in previous
interaction events.
}

\item{
Roughly twenty star cluster candidates have absolute magnitudes brighter
than $-12.0$.  The colors of these objects range from $0.0 < V-I < 1.5$,
with nine concentrated around $V-I \sim 0.6$.  For solar metallicity, a
cluster with $V-I = 0.6$ must be quite massive ($>2\times10^5$~{\msun})
 to be so luminous.  Clusters redder than $V-I = 1.0$ would have masses
$>10^7$~{\msun} which seems implausible for a single cluster; these objects
may be nucleated dwarfs or foreground stars.  However, even if
these red objects are eliminated from the luminosity function shown in
Figure~\ref{fig:lf-fig}, there are still many quite luminous clusters
remaining.  In comparison, there are no globular clusters in the Milky Way
brighter than $M_V = -11.0$ \citep{reed88}.
}
\end{enumerate}

The population of star cluster candidates 
found in Stephan's Quintet can be compared to those found in the central regions 
of traditional merger
remnants.  Schweizer \etal (1996\nocite{schweizer96}) found that the mean
colors of various star cluster systems are related to the stage of the
interaction.  Of the several mergers considered in that work, NGC~4038/39 was
found to have the bluest/youngest cluster system with median
$V-I \sim 0.4$, while NGC~7252 and NGC~3921 have few very blue clusters
and a median age consistent with the peak interaction epoch.
NGC~3256 is another example of a merger with a very young ($<20$~Myr)
star cluster system \citep{zepf99}.  It was suggested that the 
formation of new stars requires some ongoing processes such as tidal
forces or shocks.  The star cluster systems
of these different mergers also differ from each other in the
number of extremely luminous clusters; of the galaxies just discussed
NGC~7252 has a significantly larger number of $M_V < -12$ clusters.
It was hypothesized that the difference in the maximum cluster mass
could be related to the gas supply, i.e., dependent on whether one or both
of the merging galaxies were gas-rich \citep{schweizer96}.

The colors and luminosity functions of the population of clusters in Stephan's 
Quintet resembles those in some merging galaxy pairs.
This is especially true if we separately consider
the different regions of the group.  The northern starburst and
NGC~7318B/A regions have dominant young populations as do active
mergers such as the Antennae.  The tail and spiral arms of
NGC~7319 host an intermediate-aged population similar to that found
in somewhat older merger remnants.  This is consistent with the idea that
recent interaction activity of some sort is needed to generate
active star cluster formation.  However, there are clearly some clusters in
several regions of SQ which seem younger than any identified
processes.  We shall return to this point subsequently.

Despite the similarities in populations, there is a critical
difference in the spatial distribution of the cluster
population of SQ and those of merging pairs. 
The cluster populations of older merging pairs are heavily concentrated in
the inner few kpc of the galaxies.  In Stephan's Quintet, we see very few
cluster candidates near any of the galaxy nuclei, though some are found at
distances comparable to young clusters in the youngest of the Toomre
mergers, the Antennae, at 3--10~kpc.        
Though higher background
in these regions makes us less sensitive to star clusters, 
an abundant, luminous population would have been detected.
Instead, we see objects
spread through the spiral arms, the tidal debris, and even in
intergalactic regions.  This difference is likely to be due to
the fact that no actual mergers have occurred in the SQ
region within the past billion years.  Instead, SQ
has hosted high velocity encounters such as NGC~7318B with the intragroup medium
and possibly NGC~7320C with NGC~7319.
The formation of a central star cluster system must happen
as a later stage of a galaxy merger.  As NGC~7317 and NGC~7318A
show elliptical morphology, one can speculate that they 
are the products of past mergers, leading to bimodel distributions
of cluster ages.  NGC~7317 was not covered in our WFPC2 images.
NGC~7318A shows no evidence for a young nuclear population;
but an older population would likely have faded out of view in
our images.  (The Milky Way globular clusters have a peak luminosity
of $M_V \sim -10$.)

We have stated that Stephan's Quintet has few young star clusters
near the centers of the galaxies.  Alternatively, we should consider
whether galaxy pairs have young star clusters in their tidal debris
and tails such as NGC~7319.
Of the four merging pairs, Knierman \etal (2001)
found only one, NGC~3256, had a substantial population of clusters in
its tidal tails.  Because this merger was the only one of the three
without prominent tidal dwarf galaxies in the debris, they suggested
that structure preferentially formed at a certain scale.
Either dwarf galaxies or star clusters formed, but
not both.  The long tidal tail attached to NGC~7319 in Stephan's
Quintet appears to contradict this suggestion.  SQ
has a larger number of tidal dwarf candidates than any of the
other $55$ Hickson compact groups surveyed by 
Hunsberger \etal (1996)\nocite{sally96}.
Thirteen of these tidal dwarfs are in the NGC~7319 tail, which also
has a comparable number of cluster candidates.
We are to conclude that: (1) the Knierman \etal suggestion of a
preferred scale for structure formation is incorrect and based on
a coincidence in a small sample; (2) the situation is not the
same in compact groups of galaxies as it is for isolated
galaxy pairs that are merging; or (3) two separate interaction
events triggered the formation of different populations of objects
in the NGC~7319 tail.  Without more data we cannot determine which
of these ideas is correct, but in the complex compact group
environment, multiple perturbing events are likely.

Compact groups are excellent laboratories for the study of star
formation in a dynamic environment.  The relics of such interactions
might be expected to be separate generations of star clusters.
In Stephan's Quintet we certainly see evidence for several distinct
populations.  The coeval star formation is spread over surprisingly
large scales, tens of kpc, and occurs both in regions associated
with the galaxies and in the intragroup gas.  Is the formation of
star clusters in debris and outside of galaxies common in the
compact group environment?  Only one other compact group, HCG~31,
has been studied in detail with \hst/WFPC2.  There are $443$ star
clusters in the merging galaxy system, NGC~1741, in this group
\citep{johnson99}.  The median age of these clusters is extremely
young, only $4$~Myr, and there is not a significant population of
clusters older than $10$~Myr.  Unlike Stephan's Quintet, HCG~31
does host an active merging system and therefore it is not surprising
that it does have a centrally concentrated young star cluster system.
What is most striking is that the NGC~1741 system has been caught in
its infancy.

HCG~31 also has very recent ($<4$~Myr) star formation in two of the
dwarf galaxies at the redshift of the group \citep{johnson00}.
Galaxy ``E'' has both a young star cluster population and an underlying
old stellar population, similar to the large merging system, NGC~1741.
However, Galaxy ``F'' has only a young stellar population, both diffuse and 
concentrated in star clusters, and no underlying old population.
It appears to be a galaxy that formed from stripped gas out in the
tidal debris.  Johnson \etal (2000\nocite{johnson00}) point out that
a mechanism is needed to generate coeval star formation over several
galaxies in the group that are spread over $50$~kpc.  They discuss
that a shock would not propagate at a fast enough rate to explain
this scale, and raise the possibility of star formation
resulting from a series of high velocity collisions of flows of
gas, followed by delayed star formation in a small-scale cooling flow.
Stephan's Quintet requires a similar mechanism for its widespread
star formation, although it is not clear if the packaging is quite
the same; in HCG~31 the star clusters are concentrated in dwarf
galaxies.  It is worth noting that HGC~31 has only a $66$~{\kms}
velocity dispersion for its members while HCG~92, with its high
velocity encounter, has a velocity dispersion of $389$~{\kms}.

Despite this difference,
in the two cases of compact groups that have been studied, there is
evidence for recent star formation distributed over scales of tens
of kpc.  However, HCG~92 and HCG~31 are not likely to be representative
of Hickson compact groups.  They are among the few groups with
the most obvious signs of tidal interaction.  It is likely, however,
that other compact groups, particularly elliptical-rich groups,
would have had such star-forming events in the past.  This should
be recorded in the age spectrum of their cluster populations.
We might expect to see clusters out in the intragroup medium
as well as concentrated around the nuclear regions of the ellipticals.
Given the high velocity dispersion of Stephan's Quintet, it may be an 
unusual compact group.  Investigating the connection between dispersion
within a compact group and star cluster populations could offer insight 
into the mechanism for star cluster formation in the intragroup medium.
If the last interaction was more than $1$ billion years ago, even the
brightest clusters in a group such as Stephan's Quintet would have faded to
$M_V \sim -10$.  Despite the fact that deep images would be needed
to detect such faded clusters, it would be worthwhile to study some
``normal'' compact groups because of the implications that this
might have for the enrichment of the intragroup and intergalactic
medium at high redshift.  Because mergers were quite likely to be
more common at high redshift, the formation of clusters at large
distances from galaxies could allow measurable local pollution of
intergalactic gas, even by Fe-group elements that cannot be
ejected to large distances because of their Type Ia supernova origin.

\acknowledgements
Special thanks go to A. Kundu for sharing wisdom on
identification and analysis of point sources in WFPC2 images.  The thoughtful
consideration of the anonymous referee has improved this paper. This work
was supported by Space Telescope Science Institute under Grant
GO--06596.01, and by the National Science Foundation under Grant AST--0071223.
S. Gallagher also received support from NASA GSRP Grant NGT5-50277 and
from the Pennsylvania Space Grant Consortium.

%%%%%%%%%%%%%%%%%%%%%%%%%%%%%%%%%%%%%%%%%%%%%%%%%%%%%%%%%%%%%%%%%%%%%%%%%%%%%%%%%%

\clearpage
%%%%%%%%%%%%%%%%%%%%%%%%%%%%%%%%%%%%%%%%%%%%%%%%%%%%%%%%%%%%%%%%%%%%%%%%%%%%%%%%%

\begin{figure*}[t]
\figurenum{1}
%\centerline{\includegraphics[scale=0.5,angle=-90]{fig1.ps}}
\protect\caption{$R-$band image of Stephan's Quintet taken with the 1.5~m Palomar 
telescope.  North is up and east is to the left.  The 
members of HCG~92 have been labeled in addition to NGC~7320C and specific
regions of interest.  NGC~7320 is a foreground galaxy. Adapted from 
Hunsberger, Charlton \& Zaritsky (1996).  The $3^{\prime}.7\times2^{\prime}.5$
region imaged with WFPC2 in two pointings is outlined with the gray box.
\label{fig:palomar-fig} 
}
\end{figure*}

%\clearpage

\begin{figure*}[t]
\figurenum{2}
%\centerline{\includegraphics[scale=0.5,angle=-90]{fig2.ps}}
\protect\caption
{Composite WFPC2 $V-$band image of Stephan's Quintet with the star cluster
candidates indicated with open circles.  The display scale is logarithmic, and the 
pixels have been binned by a factor of two in $x$ and $y$.  
These point sources were found in all $B$, $V$, and $I$ images and have color 
errors less than 0.2 mag.  The regions corresponding to the symbols in 
Figures~\ref{fig:cc-fig} and ~\ref{fig:cmd-fig} have been
labeled.  Note that the immediate vicinity of the foreground galaxy NGC 
7320 was excluded from the source searching algorithm.
\label{fig:sq-all-hst}
}
\end{figure*}

%\clearpage

\begin{figure*}[t]
\figurenum{3}
%\centerline{\includegraphics[scale=0.5,angle=-90]{fig3a.ps}}
\protect\caption
{{\bf (a)} Cropped \hst\ $V-$band image centered on NGC~7319 and 7318B/A. 
White contours indicate background values of 7, 10, 20, 50, 100, and 200 DN.  That 
part of the image not shown has background values $<7$~DN except near bright point sources.
\label{fig:completeness}
}
\end{figure*}

\begin{figure*}[t]
\figurenum{3}
\plotone{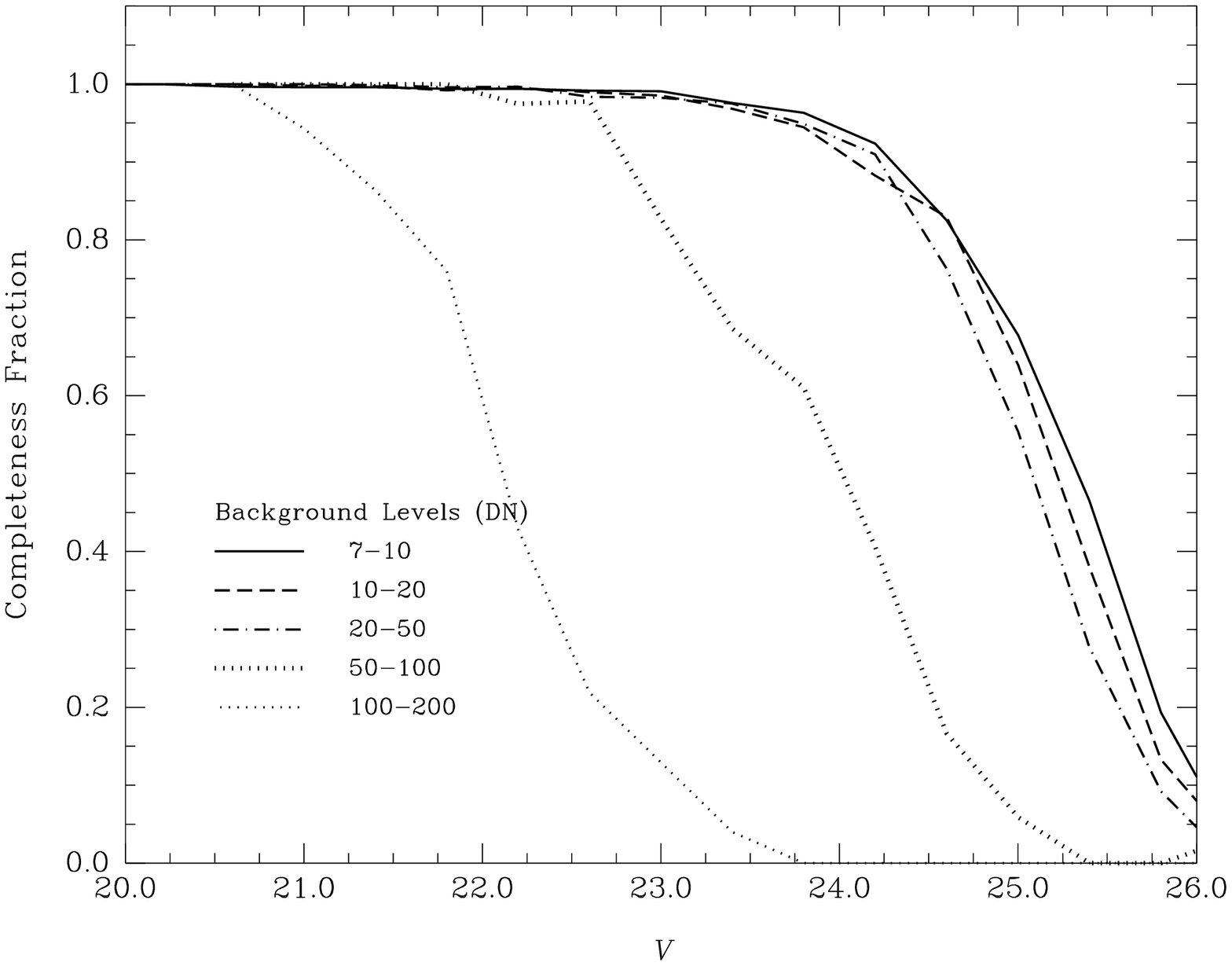}
\protect\caption{
{\bf (b)} Completeness fraction versus apparent $V$ magnitude for point sources.  
For background values $<50$~DN, the sample is $\gtrsim90\%$ complete at 24.25 mag and 
$\gtrsim50\%$ complete at 25.1 mag. 
\label{fig:completeness}
}
\end{figure*}

\begin{figure*}[t]
\figurenum{4}
\plotone{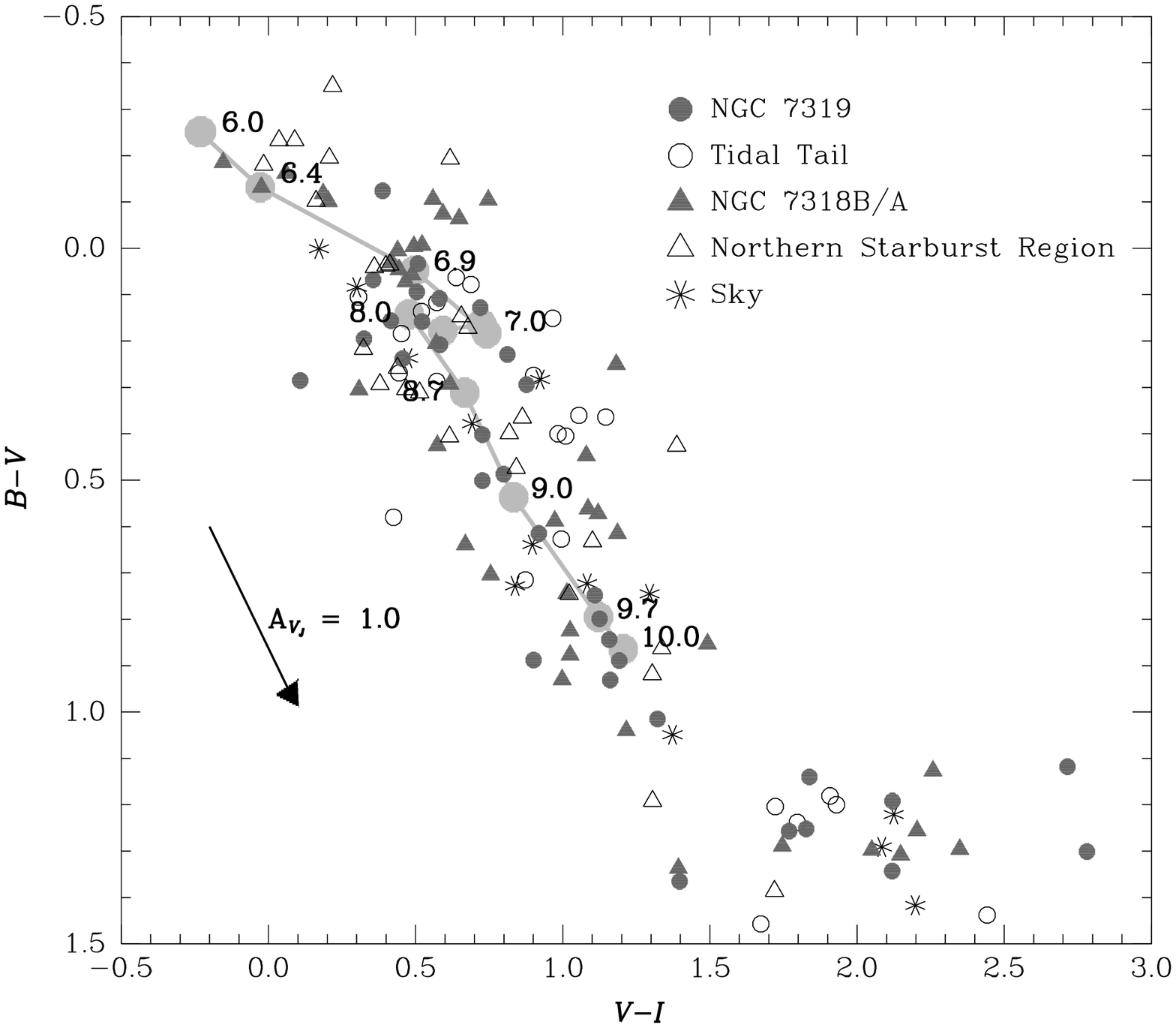}
\protect\caption{$B-V$ versus $V-I$ color-color diagram of the point sources in Stephan's 
Quintet with color errors less than 0.2 mag.  
The thick, gray line represents the evolutionary track derived from the 
1995 Bruzual \& Charlot (1993) instantaneous-burst stellar population synthesis models (with
a Salpeter IMF and solar metallicity).  Labels along the track indicate the logarithm of 
the age in years.  The photometry of the point sources has been corrected for
Galactic extinction of $E(B-V)$=0.083 (Burstein \& Heiles 1984).
The symbols, as indicated in the legend, represent the physical regions as defined in 
Figure~{\ref{fig:sq-all-hst}}.  The reddening vector, $A_{V_J}$, represents one magnitude
of extinction in the Johnson (rather than \hst) $V-$band filter ($E(B-V)=0.32$). 
\label{fig:cc-fig}
}
\end{figure*}

%\clearpage

\begin{figure*}[t]
\figurenum{5}
\plotone{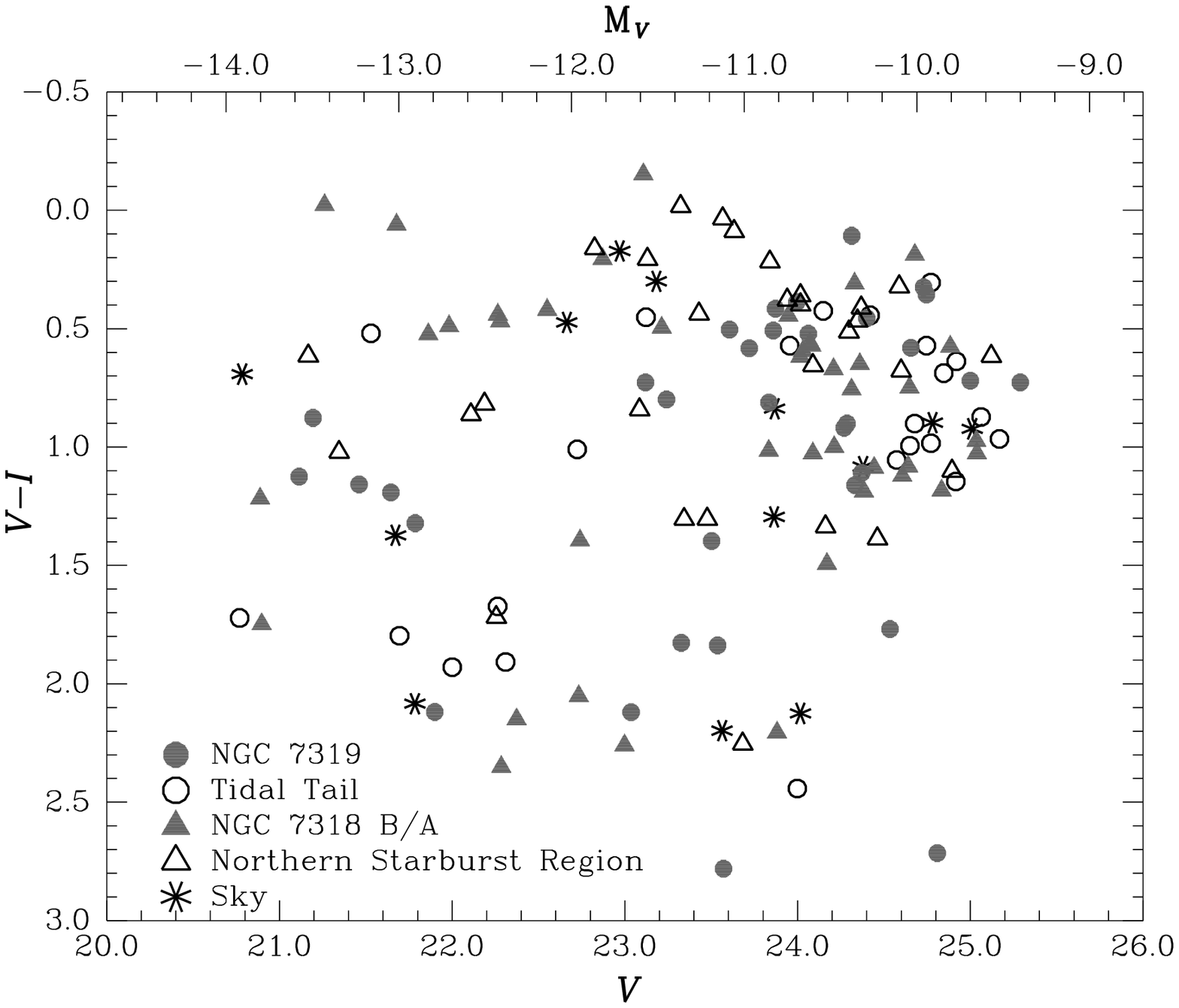}
\protect\caption{$V-I$ versus $V$ color-magnitude diagram of the point sources in 
Stephan's Quintet with color errors less than 0.2 mag.  Objects with $V-I>1.5$ are
likely to be stars. $M_V$ was calculated assuming a redshift distance modulus of 
34.69 corresponding to 85~Mpc for $H_0=75$~km~s$^{-1}$~Mpc$^{-1}$ and $q_0=0.1$.  
Symbols as in Figure~{\ref{fig:cc-fig}}.
\label{fig:cmd-fig}
}
\end{figure*}

%\clearpage

\begin{figure*}[t]
\figurenum{6}
\plottwo{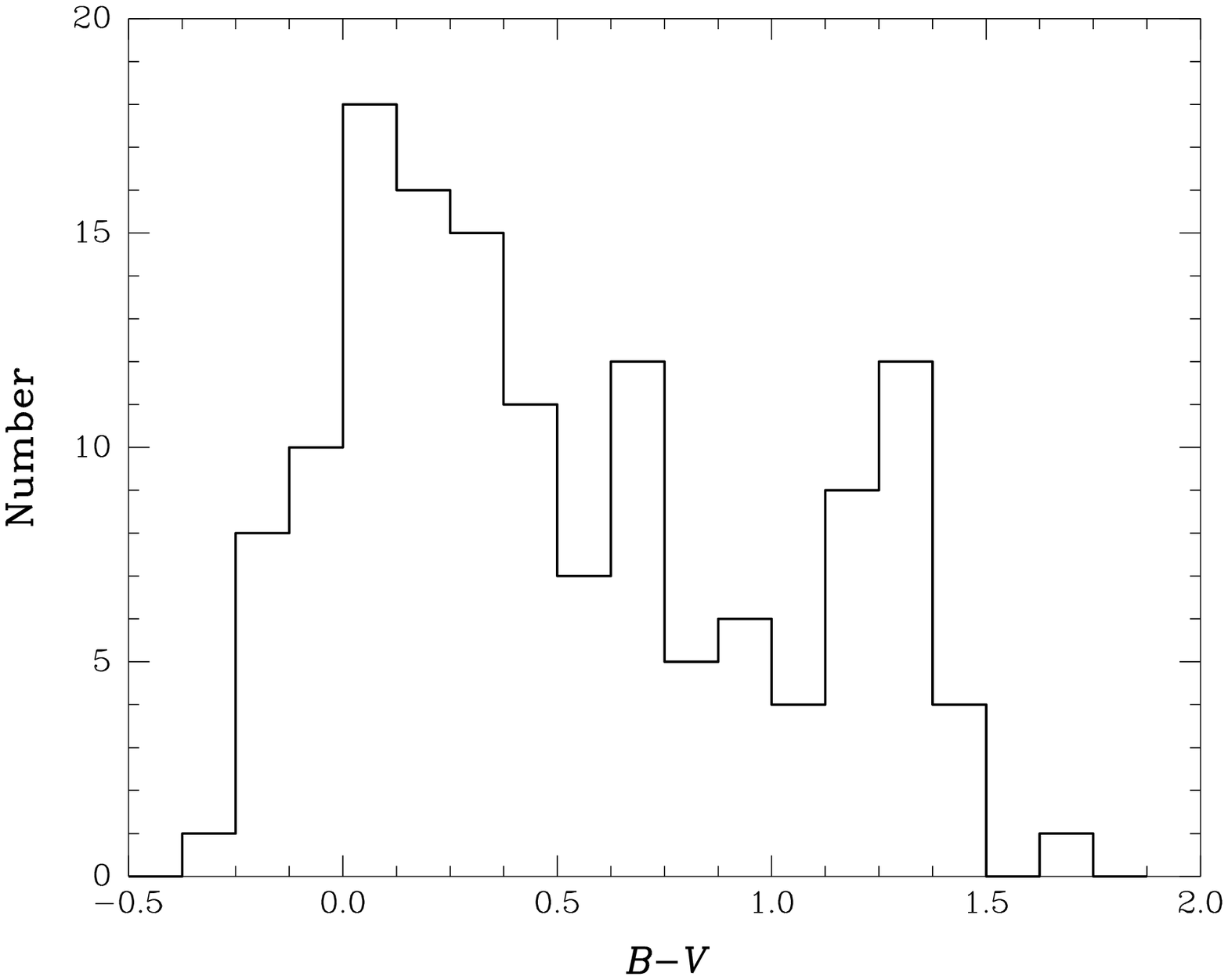}{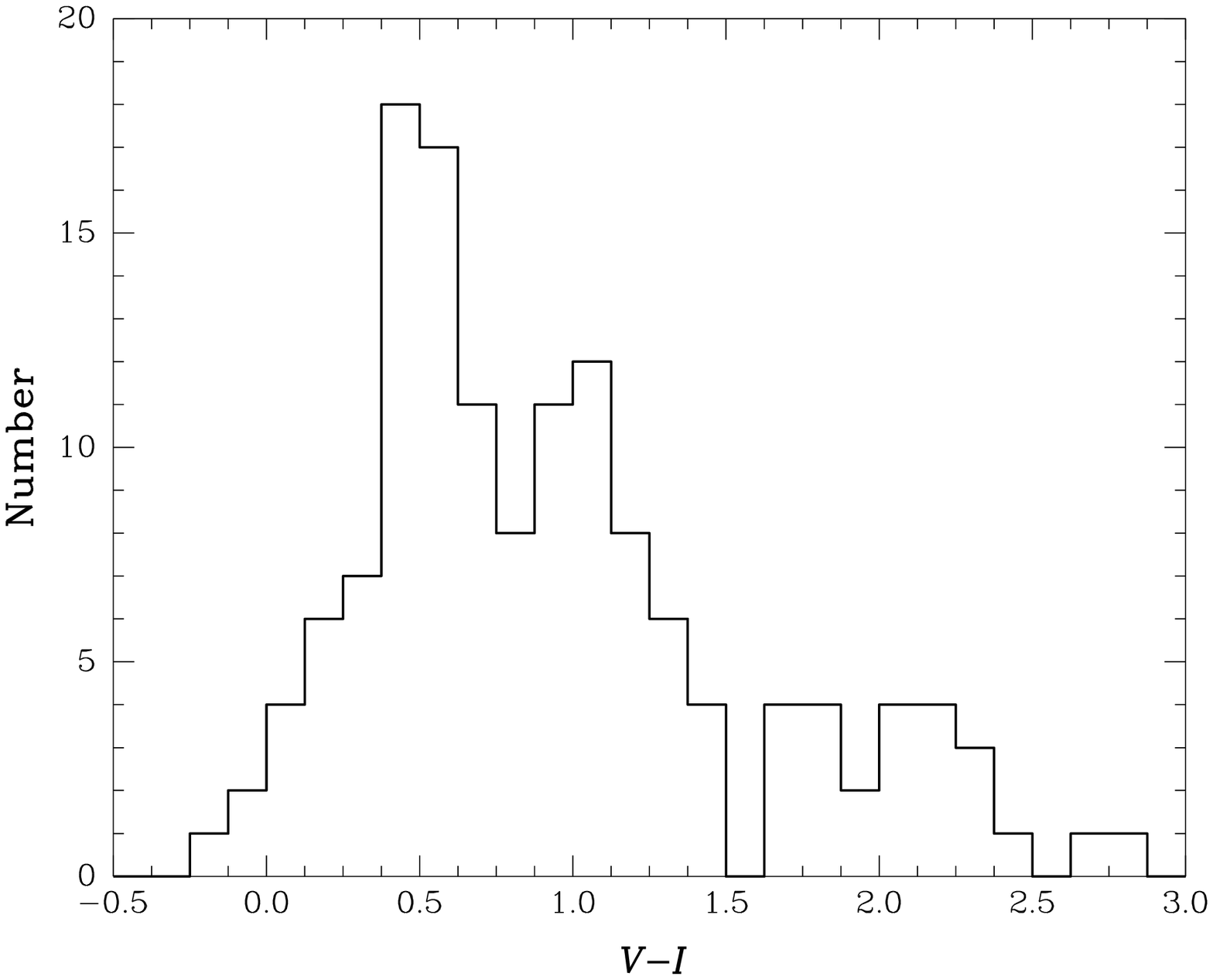}
\protect\caption{Histograms of the $V-I$ and $B-V$ colors for all point
sources found in the field.  Note the inflection in the right-hand histogram at $V-I\sim1.5$
corresponding to distinct star cluster candidate and stellar populations.
\label{fig:colorhis}
}
\end{figure*}

%\clearpage

\begin{figure*}[t]
\figurenum{7}
\plotone{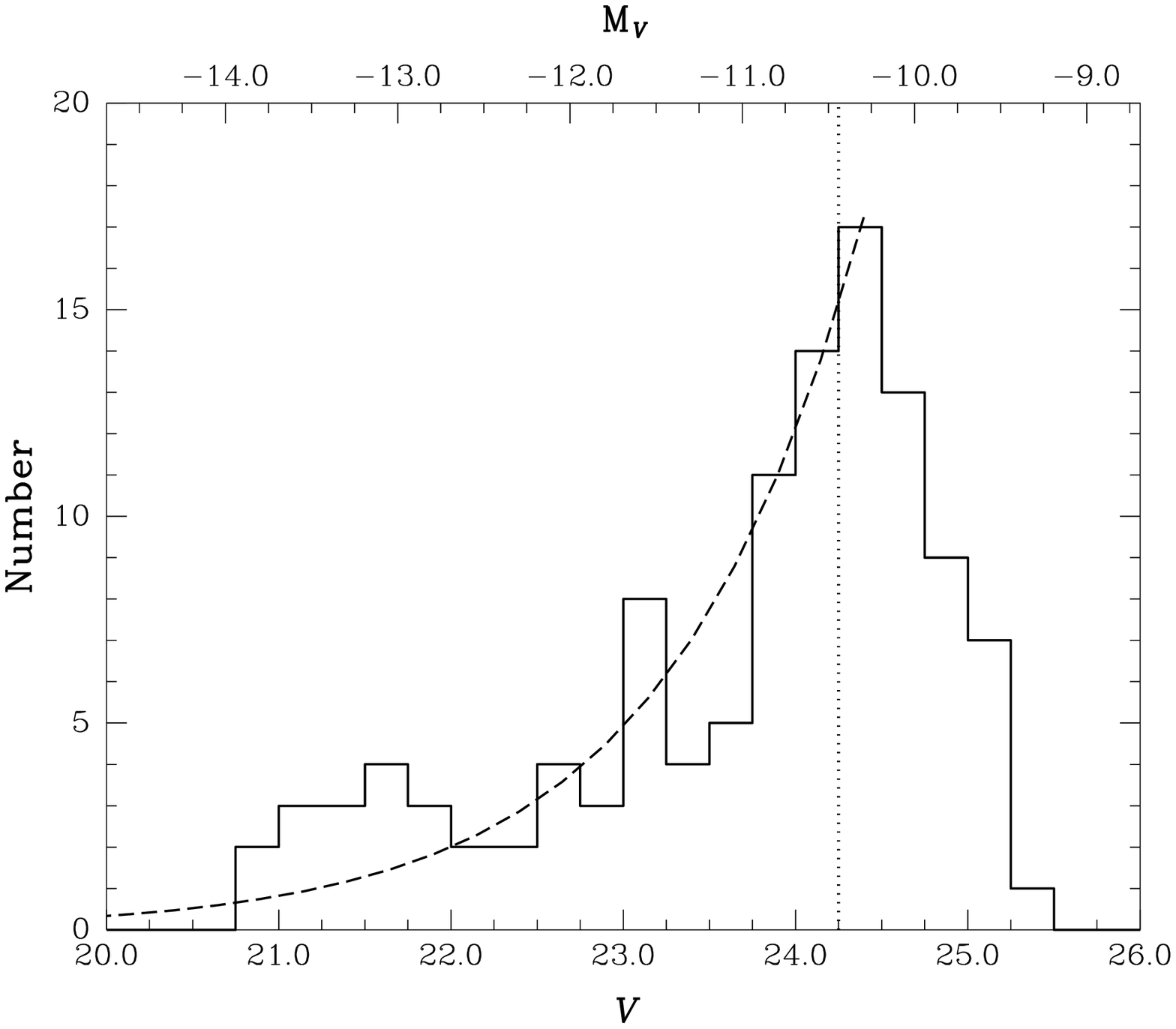}
\protect\caption{The $V$--band luminosity function for all candidate star
clusters with $V-I<1.5$.  The vertical dotted line represents the $\gtrsim90\%$ completeness
limit for all sources with background values $<50$~DN.  
The absolute magnitude scale was determined as in Figure~{\ref{fig:cmd-fig}}.
\label{fig:lf-fig}
}
\end{figure*}

%\clearpage

\begin{figure*}[t]
\figurenum{8}
\plottwo{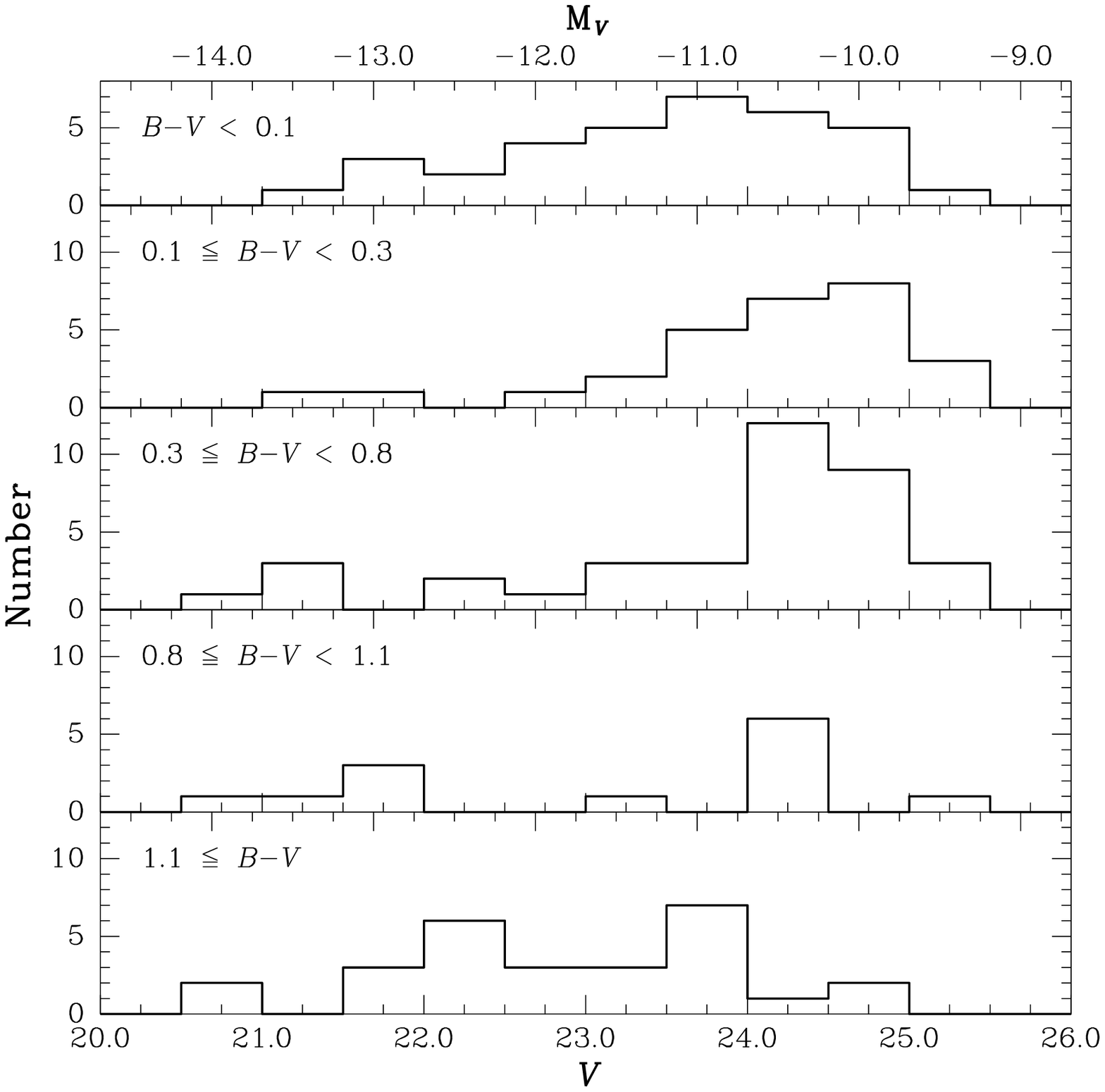}{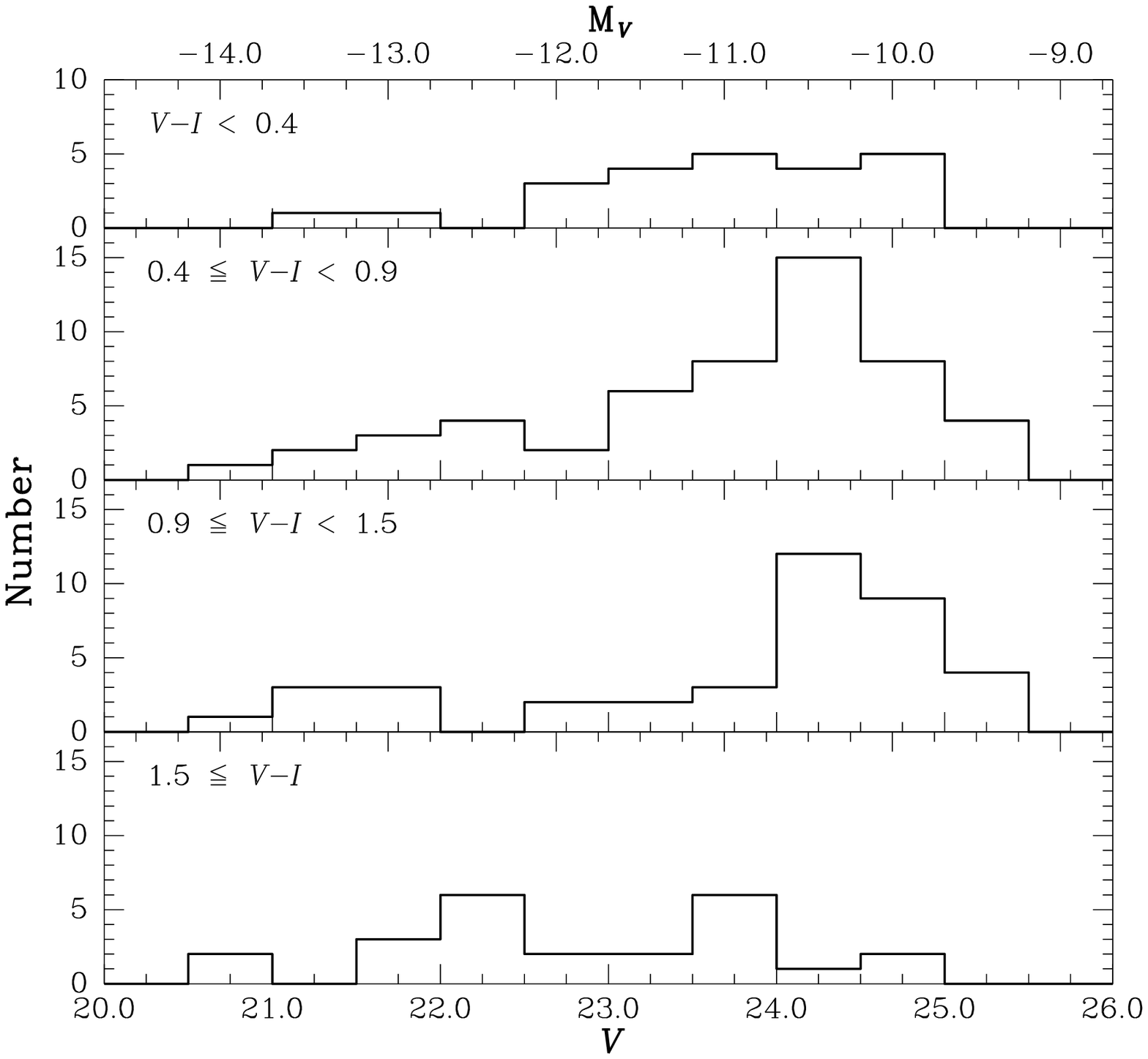}
\protect\caption
{Histograms of the point sources found in Stephan's Quintet per $V$ magnitude bin
divided into color groups as labeled.  $M_V$ was calculated
as in Figure~{\ref{fig:cmd-fig}}.  
{\bf(a)} These groups correspond to the $B-V$ color-groups 
A, B, C, D, and S first defined in the caption to Figure~{\ref{fig:tail-figs}}a.
{\bf(b)} $V-I$ color groups defined as in Schweizer et al. (1996).
\label{fig:histograms}
}
\end{figure*}

%\clearpage

\begin{figure*}[t]
\figurenum{9}
%\centerline{\includegraphics[scale=0.45,angle=-90]{fig9a.ps}}
\protect\caption
{{\bf(a)} Close-up $V-$band image of the tidal tail region in Stephan's Quintet.
This image is displayed as in Figure~{\ref{fig:sq-all-hst}} although 
it extends farther to the west than the corresponding region in that figure.  
Large, open circles mark the locations
of star cluster candidates considered to be in the tidal tail.  The other symbols
represent $B-V$ color groups: 
Group A, $B-V<0.1$, ($\triangle$); 
Group B, $0.1\le B-V<0.3$, ($\Box$); 
Group C, $0.3\le B-V<0.8$, (open pentagon);
Group D, $0.8\le B-V<1.1$, ($\bigcirc$); and
Group S, $1.1\ge B-V$, (open star).
The number labels are for ease of identification with the source list in 
Table~1.
\label{fig:tail-figs}
}
\end{figure*}

\begin{figure*}[t]
\figurenum{9}
\plotone{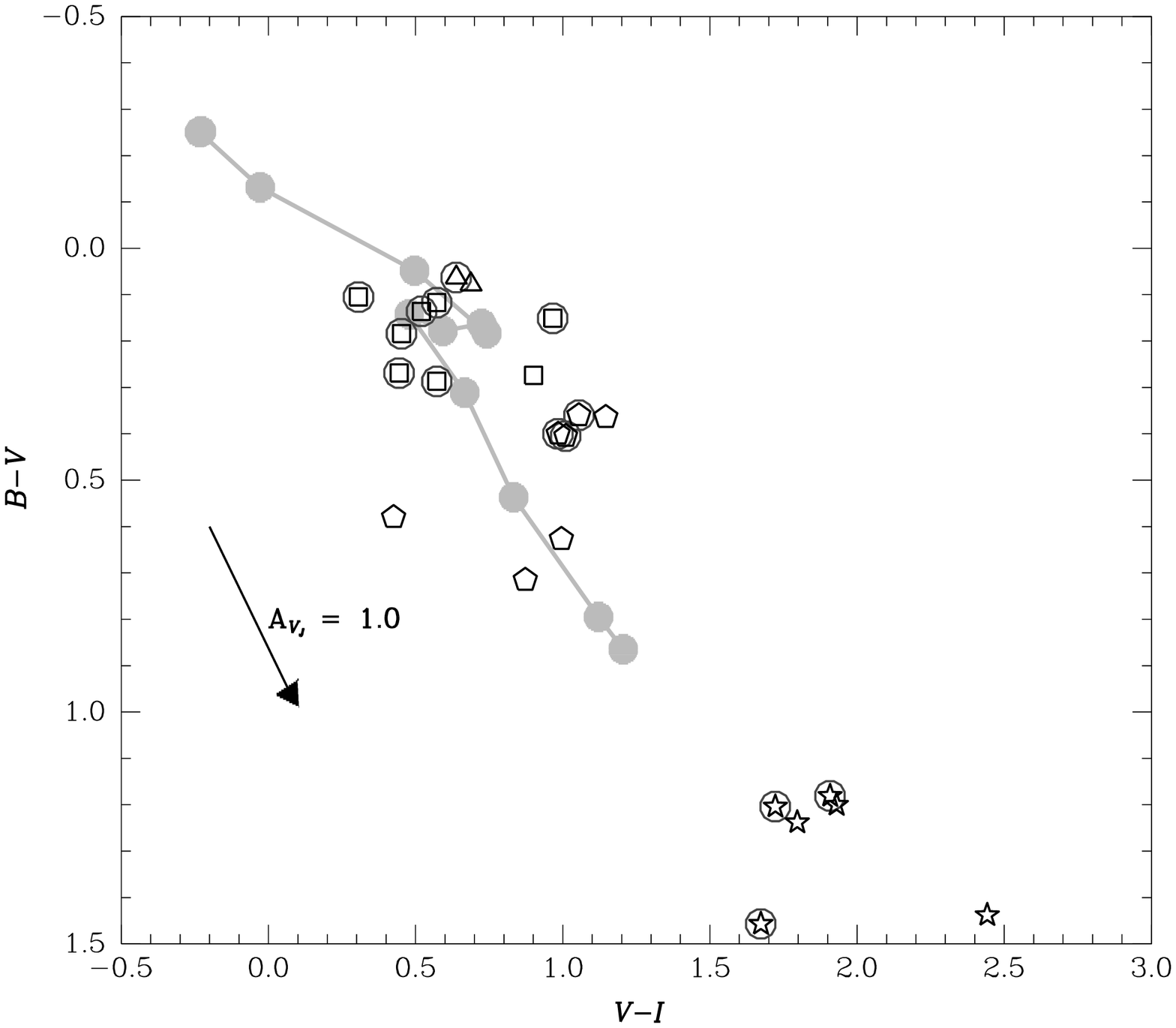}
\protect\caption{
{\bf(b)} Color-color diagram of the point sources in the tidal tail region.  
The thick, gray line represents the evolutionary track derived from the 
1995 Bruzual \& Charlot (1993) instantaneous-burst stellar population synthesis model (with
a Salpeter IMF and solar metallicity).  Large, filled circles along the track correspond 
to the ages as labeled in Figure~{\ref{fig:cc-fig}}.  A reddening 
vector, $A_{V_J}$, corresponding to one magnitude of extinction in the Johnson $V$ 
filter ($E(B-V)=0.32$) is included for reference.
\label{fig:tail-figs}
}
\end{figure*}

%\clearpage

\begin{figure*}[t]
\figurenum{10}
%\plotone{fig10a.ps}
\protect\caption
{{\bf(a)} Close-up $V-$band image of the sky region in Stephan's Quintet
displayed as in Figure~{\ref{fig:sq-all-hst}}. The symbols
represent $B-V$ color groups as defined in the caption of Figure~{\ref{fig:tail-figs}},
and the number
labels are for ease of identification with the source list in Table~2.
\label{fig:sky-figs}
}
\end{figure*}

\begin{figure*}[t]
\figurenum{10}
\plotone{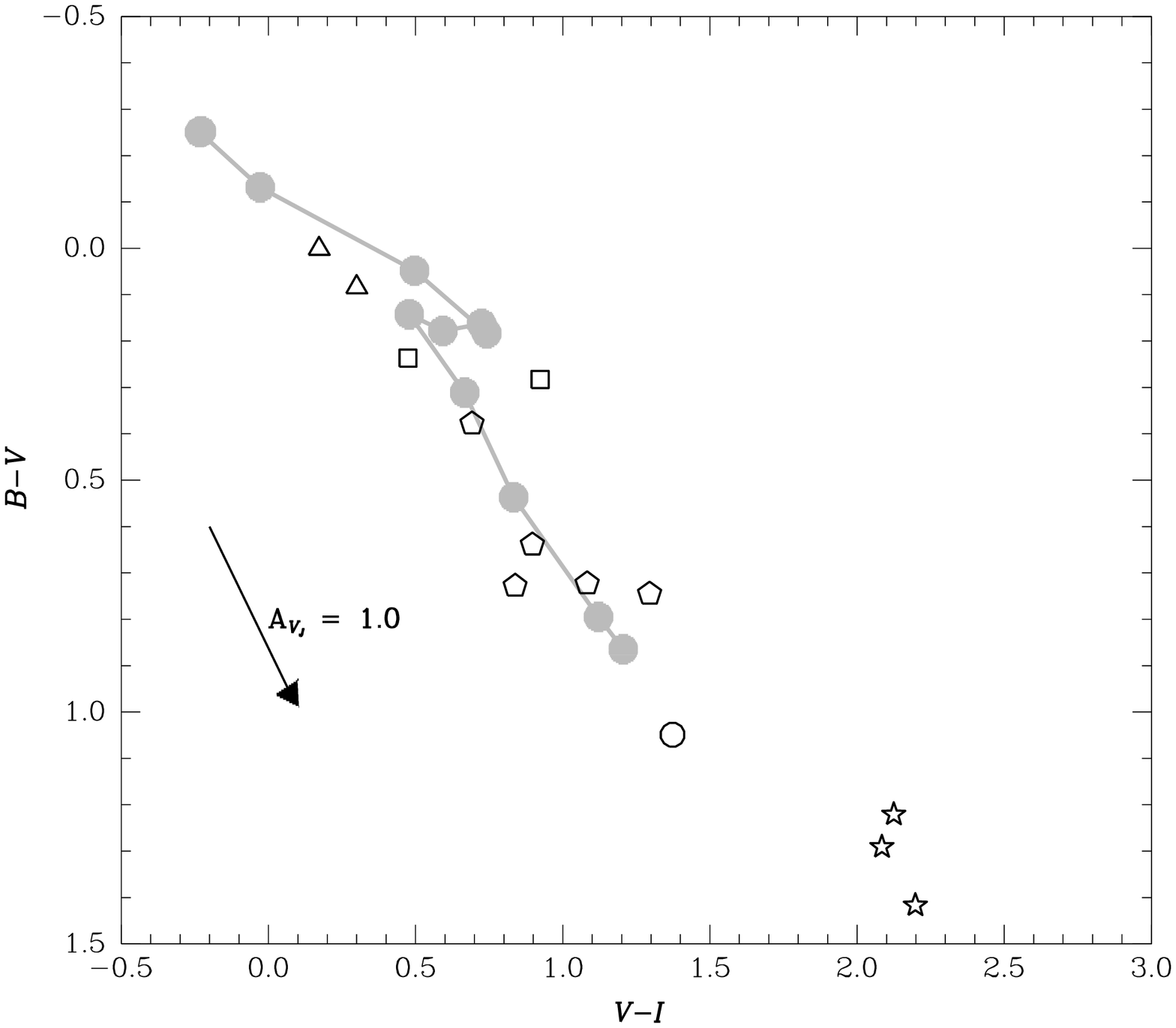}
\protect\caption{
{\bf(b)} Color-color diagram of the point sources in the sky region. 
The thick, gray line represents the evolutionary track derived from the 
1995 Bruzual \& Charlot (1993) instantaneous-burst stellar population synthesis model (with
a Salpeter IMF and solar metallicity).  Large, filled circles along the track correspond 
to the ages as labeled in Figure~{\ref{fig:cc-fig}}. 
A reddening vector, $A_{V_J}$, 
corresponding to one magnitude of extinction in the Johnson $V$ filter ($E(B-V)=0.32$)
is included for reference. Symbols are as defined in Figure~{\ref{fig:tail-figs}}.
\label{fig:sky-figs}
}
\end{figure*}

%\clearpage

\begin{figure*}[t]
\figurenum{11}
%\plotone{fig11a.ps}
\protect\caption
{{\bf(a)} Close-up $V-$band image of NGC~7319 displayed as in Figure~{\ref{fig:sq-all-hst}}. 
The symbols represent $B-V$ color groups as defined in the caption of
Figure~{\ref{fig:tail-figs}}, and the number
labels are for ease of identification with the source list in Table~3.
Note that sources are listed in the table corresponding to the image in which they first
appear.
\label{fig:7319-panel}
}
\end{figure*}

\begin{figure*}[t]
\figurenum{11}
\plotone{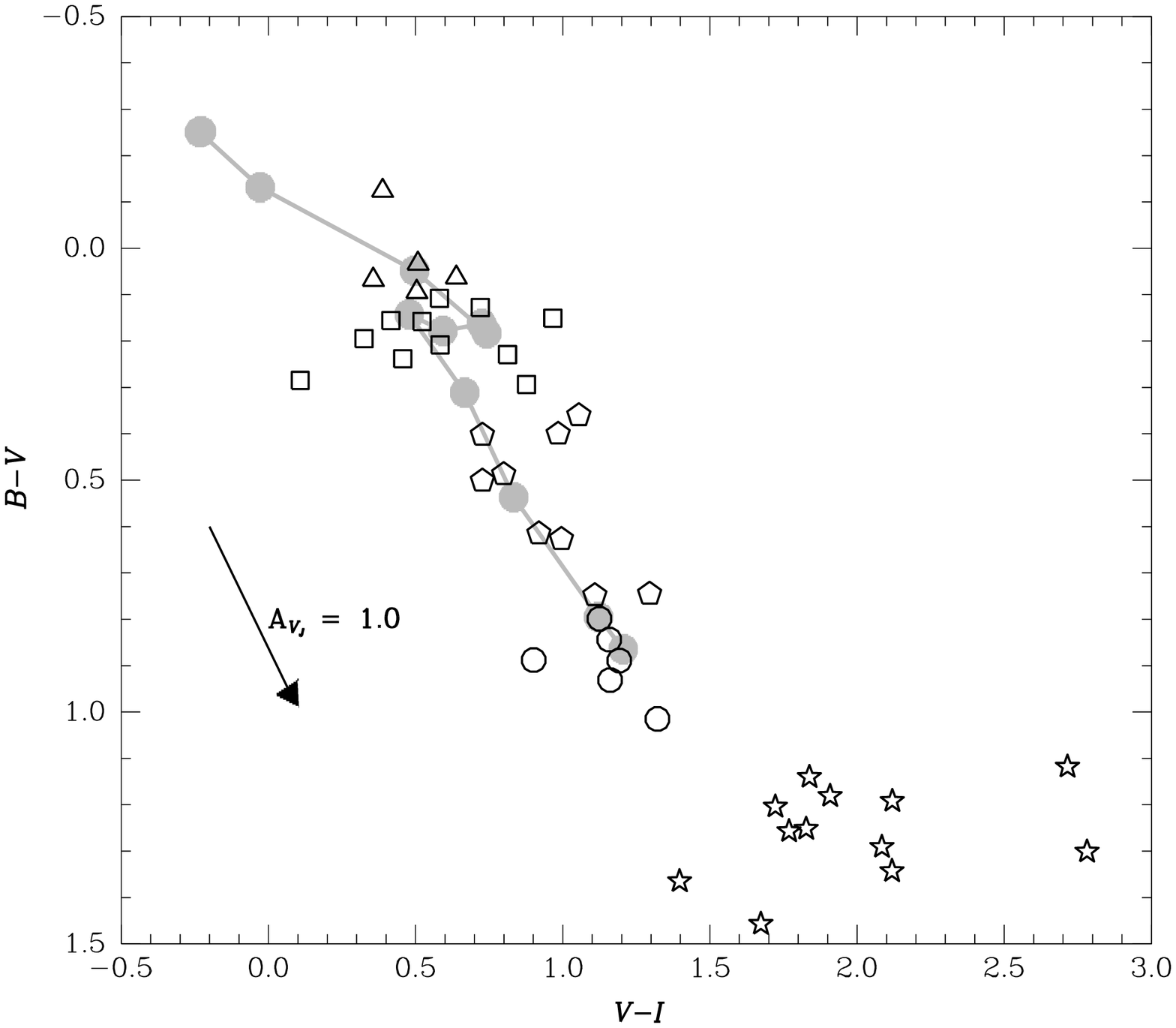}
\protect\caption{
{\bf (b)} Color-color diagram of the point sources in the vicinity of NGC~7319.  
The thick, gray line represents the evolutionary track derived from the 
1995 Bruzual \& Charlot (1993) instantaneous-burst stellar population synthesis model (with
a Salpeter IMF and solar metallicity).  Large, filled circles along the track correspond 
to the ages as labeled in Figure~{\ref{fig:cc-fig}}.
A reddening vector, $A_{V_J}$, 
corresponding to one magnitude of extinction in the Johnson 
$V$ filter ($E(B-V)=0.32$) is included for reference.
Symbols are as defined in Figure~{\ref{fig:tail-figs}}. 
\label{fig:7319-cc}
}
\end{figure*}

%\clearpage

\begin{figure*}[t]
\figurenum{12}
%\plotone{fig12a.ps}
\protect\caption
{{\bf(a)} Close-up $V-$band image of NGC~7318 and the northern starburst region
displayed as in Figure~{\ref{fig:sq-all-hst}}. 
The symbols represent $B-V$ color groups as defined in the caption of
Figure~{\ref{fig:tail-figs}}, and the number
labels are for ease of identification with the source list in Table~4.
Note that sources are listed in the table corresponding to the image in which they first
appear.
\label{fig:7318-panel}
}
\end{figure*}

\begin{figure*}[t]
\figurenum{12}
\plotone{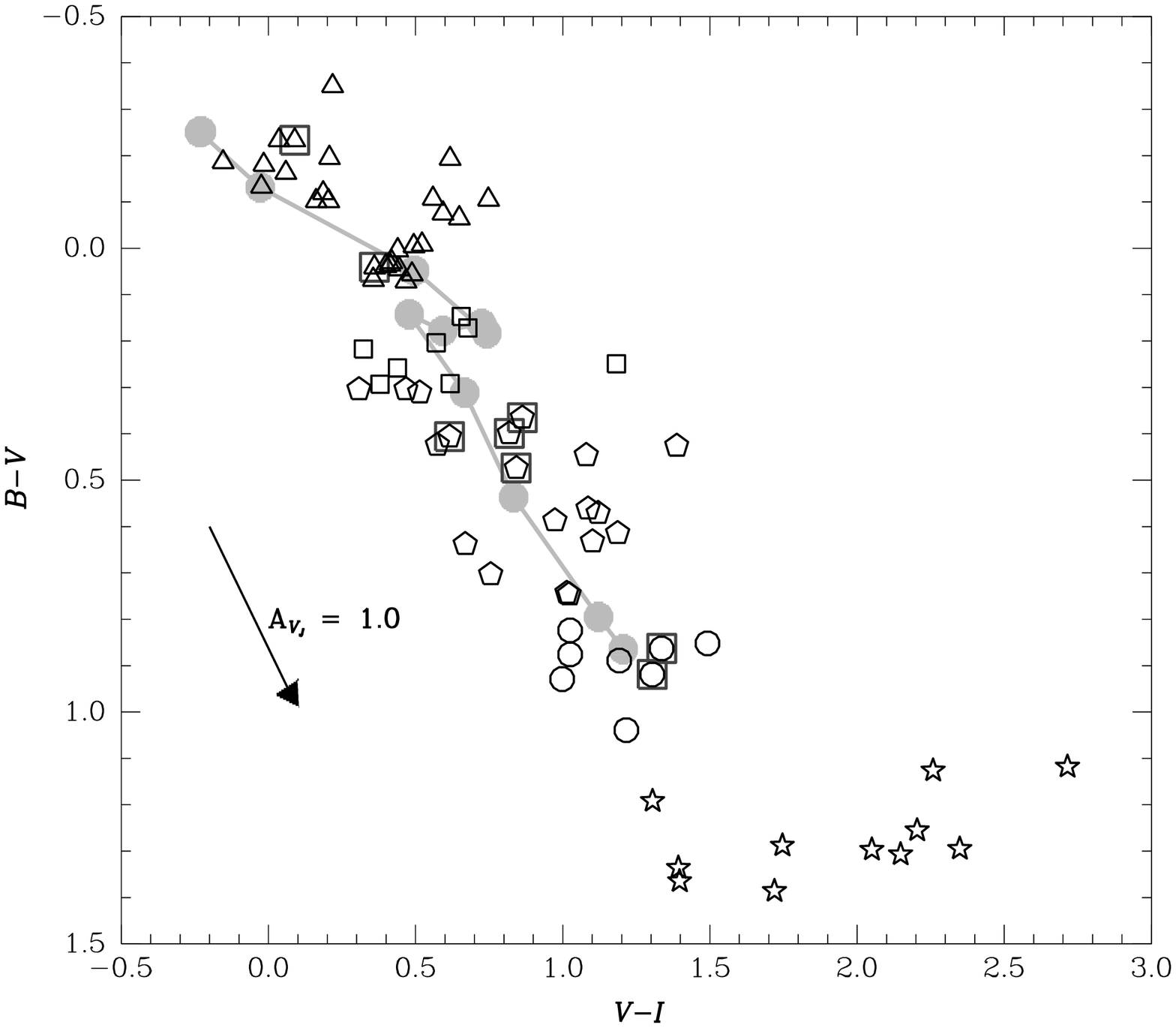}
\protect\caption{
{\bf (b)} Color-color diagram of the point sources in the vicinity of NGC~7319.  
The thick, gray line represents the evolutionary track derived from the 
1995 Bruzual \& Charlot (1993) instantaneous-burst stellar population synthesis model (with
a Salpeter IMF and solar metallicity).  Large, filled circles along the track correspond 
to the ages as labeled in Figure~{\ref{fig:cc-fig}}.
A reddening vector, $A_{V_J}$, 
corresponding to one magnitude of extinction in the Johnson 
$V$ filter ($E(B-V)=0.32$) is included for reference.
Symbols are as defined in Figure~{\ref{fig:tail-figs}}; the point sources associated with the
dwarf galaxy (see Figure~{\ref{fig:dwarf-fig}}) are boxed. 
\label{fig:7318-cc}
}
\end{figure*}

%\clearpage

\begin{figure*}[t]
\figurenum{13}
\plotone{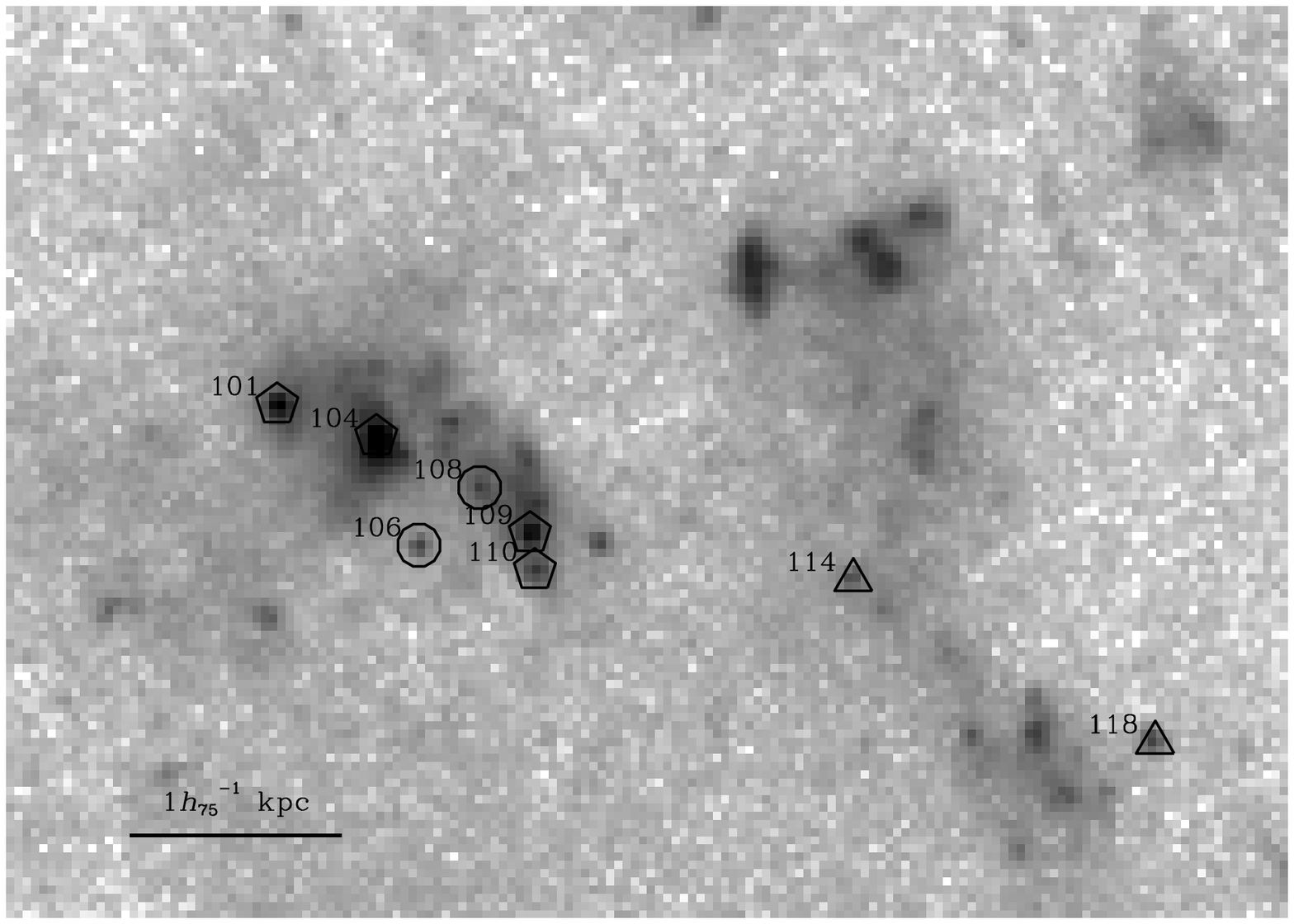}
\protect\caption{
Close-up $V-$band image of the dwarf galaxy in the northern starburst region displayed on a logarithmic scale at full resolution. The symbols represent $B-V$ color groups as defined in Figure~{\ref{fig:tail-figs}}, and the number labels are for ease of identification with the source list in
Table~4. These sources are boxed in Figure~{\ref{fig:7318-cc}}. \label{fig:dwarf-fig}
}
\end{figure*}

\clearpage

\begin{figure*}
\figurenum{14}
%\centerline{\includegraphics[scale=0.6,angle=-90]{fig14.ps}}
%\plotone{fig14.ps}
\protect\caption{
Top: Long-slit spectrum at $R\sim600$, taken with the Marcario Low-Resolution Spectrograph on the Hobby-Eberly Telescope, of a region of tidal debris apparently related to the galaxy NGC 7318B.  The spectrum shows, from left to right, H$\beta$, the {\OIII} doublet,H$\alpha$, {\NII}, and the {\SII} doublet. Left: The location of the slit with five debris regions marked. Right: A close-up view of the region around the H$\alpha$ line showing approximate velocities for each of the debris regions. \label{fig:het}
}
\end{figure*}

\clearpage

%%%%%%%%%%%%%%%%%%%%%%%%%%%%%%%%%%%%%%%%%%%%%%%%%%%%%%%%%%%%%%%%%%%%%%%%%%%%%%%%%
%
%Table of photometric data for the tidal tail region.
\begin{deluxetable}{rrccrcccl}
\tablewidth{0pt}
\tablecaption{Photometric Data: Tidal Tail.{\tablenotemark{a}}
\label{tab:tail-table}}
\tablehead{
\colhead{Label{\tablenotemark{b}}} &
\colhead{$M_V${\tablenotemark{c}}} & 
\colhead{$V$} &
\colhead{$V_{\rm err}$} &
\colhead{$B-V$} &
\colhead{$(B-V)_{\rm err}$} &
\colhead{$V-I$} &
\colhead{$(V-I)_{\rm err}$} &
\colhead{Notes{\tablenotemark{d}}}
}
\startdata
 1 & $ -9.63 $& 25.06 & 0.13 & 0.71 & 0.19 & 0.87 & 0.17 & C\\
 2 & $-10.27 $& 24.42 & 0.10 & 0.27 & 0.13 & 0.44 & 0.16 & B, T\\
 3 & $-10.54 $& 24.15 & 0.06 & 0.58 & 0.09 & 0.42 & 0.10 & C\\
 4 & $-10.73 $& 23.95 & 0.06 & 0.12 & 0.07 & 0.57 & 0.08 & B, T\\
 5 &   ------ & 24.00 & 0.06 & 1.44 & 0.14 & 2.44 & 0.07 & S\\
 6 &   ------ & 21.70 & 0.01 & 1.24 & 0.02 & 1.80 & 0.01 & S\\
 7 & $-11.97 $& 22.72 & 0.02 & 0.41 & 0.03 & 1.01 & 0.03 & C, T\\
 8 & $ -9.84 $& 24.85 & 0.10 & 0.08 & 0.12 & 0.69 & 0.16 & A\\
 9 & $ -9.94 $& 24.75 & 0.09 & 0.29 & 0.12 & 0.57 & 0.15 & B, T\\
10 & $ -9.92 $& 24.77 & 0.11 & 0.10 & 0.15 & 0.31 & 0.19 & B, T\\
11 & $-11.57 $& 23.12 & 0.03 & 0.18 & 0.04 & 0.45 & 0.05 & B, T\\
12 & $-13.16 $& 21.53 & 0.01 & 0.14 & 0.01 & 0.52 & 0.02 & A, T\\
13 & $-10.01 $& 24.68 & 0.08 & 0.27 & 0.10 & 0.90 & 0.11 & B\\
14 & $ -9.77 $& 24.92 & 0.09 & 0.36 & 0.12 & 1.15 & 0.12 & C\\
15 &   ------ & 22.00 & 0.01 & 1.20 & 0.03 & 1.93 & 0.02 & S\\
16 &   ------ & 22.31 & 0.02 & 1.18 & 0.03 & 1.91 & 0.02 & S, T\\
17 &   ------ & 22.26 & 0.02 & 1.46 & 0.03 & 1.67 & 0.02 & S, T\\
18 &   ------ & 20.77 & 0.01 & 1.20 & 0.01 & 1.72 & 0.01 & S, T\\
19 & $-10.12 $& 24.57 & 0.08 & 0.36 & 0.10 & 1.05 & 0.10 & C, T\\
20 & $ -9.52 $& 25.17 & 0.12 & 0.15 & 0.16 & 0.97 & 0.17 & B, T\\ 
21 & $-10.04 $& 24.65 & 0.08 & 0.63 & 0.12 & 0.99 & 0.11 & C\\
22 & $ -9.77 $& 24.92 & 0.10 & 0.06 & 0.12 & 0.64 & 0.16 & A, T\\
23 & $ -9.92 $& 24.77 & 0.08 & 0.40 & 0.12 & 0.98 & 0.12 & C, T\\
\enddata
\tablenotetext{a}{All photometry has been corrected for Galactic extinction as described in 
$\S$~\ref{sec:finalphot}.}
\tablenotetext{b}{The numeric labels identify the sources as labeled in
Figure~\ref{fig:tail-figs}a.}
\tablenotetext{c}{ $M_V$ was calculated assuming a redshift distance modulus of 
34.69 corresponding to 
85~Mpc for $H_0=75$~km~s$^{-1}$~Mpc$^{-1}$ and $q_0=0.1$.  Note that the
absolute magnitudes of putative stars are marked with -----.}
\tablenotetext{d}{Letters A, B, C, D, and S 
correspond to $B-V$ color groups as defined in the 
caption to Figure~{\ref{fig:tail-figs}}.  The star cluster candidates apparently
associated with the tidal tail are designated with the letter T.}
\end{deluxetable}
%\clearpage
%--------------------------------------------------------------------------------------
%Table of photometric data for the sky region.
\begin{deluxetable}{rrccrcccl}
\tablewidth{0pt}
\tablecaption{Photometric Data: Sky Region.{\tablenotemark{a}}
\label{tab:sky-table}}
\tablehead{
\colhead{Label{\tablenotemark{b}}} &
\colhead{$M_V${\tablenotemark{c}}} & 
\colhead{$V$} &
\colhead{$V_{\rm err}$} &
\colhead{$B-V$} &
\colhead{$(B-V)_{\rm err}$} &
\colhead{$V-I$} &
\colhead{$(V-I)_{\rm err}$} &
\colhead{Notes{\tablenotemark{d}}}
}
\startdata
24 & $ -13.91 $ &  20.78 &  0.01 &  0.38 &        0.01 &  0.69 &       0.01 & C  \\
25 & $ -13.02 $ &  21.67 &  0.01 &  1.05 &        0.02 &  1.37 &       0.02 & D  \\
26 & $  -9.68 $ &  25.01 &  0.13 &  0.28 &        0.16 &  0.92 &       0.17 & B  \\
27 & $ -10.82 $ &  23.87 &  0.05 &  0.73 &        0.08 &  0.84 &       0.07 & C  \\
28 &   -----    &  23.56 &  0.04 &  1.42 &        0.09 &  2.20 &       0.04 & S  \\
29 & $ -10.31 $ &  24.38 &  0.06 &  0.72 &        0.10 &  1.08 &       0.09 & C  \\
30 & $ -11.72 $ &  22.97 &  0.03 &  0.00 &        0.04 &  0.17 &       0.06 & A  \\
31 & $ -11.51 $ &  23.18 &  0.04 &  0.08 &        0.04 &  0.30 &       0.06 & A  \\
32 & $  -9.91 $ &  24.78 &  0.09 &  0.64 &        0.13 &  0.90 &       0.12 & C  \\
33 & $ -12.03 $ &  22.66 &  0.02 &  0.24 &        0.03 &  0.47 &       0.03 & B  \\
34 &   -----    &  24.02 &  0.05 &  1.22 &        0.10 &  2.12 &       0.06 & S  \\
35 & $ -10.83 $ &  23.86 &  0.04 &  0.74 &        0.07 &  1.29 &       0.06 & C  \\
36 &   -----    &  21.78 &  0.01 &  1.29 &        0.02 &  2.08 &       0.01 & S  \\
\enddata
\tablenotetext{a}{All photometry has been corrected for Galactic extinction as described in 
$\S$~\ref{sec:finalphot}.}
\tablenotetext{b}{The numeric labels identify the sources as labeled in
Figure~\ref{fig:sky-figs}a.}
\tablenotetext{c}{ $M_V$ was calculated assuming a redshift distance modulus of 
34.69 corresponding to 
85~Mpc for $H_0=75$~km~s$^{-1}$~Mpc$^{-1}$ and $q_0=0.1$.  Note that the
absolute magnitudes of putative stars are marked with -----.}
\tablenotetext{d}{Letters correspond to $B-V$ color groups as defined in the 
caption to Figure~{\ref{fig:tail-figs}}.}
\end{deluxetable}
%
%\clearpage
%
%--------------------------------------------------------------------------------------
%Table of photometric data for the region around 7319.
\begin{deluxetable}{rrccrcccl}
\tablewidth{0pt}
\tablecaption{Photometric Data: NGC~7319 and Vicinity.{\tablenotemark{a}}
\label{tab:7319-table}}
\tablehead{
\colhead{Label{\tablenotemark{b}}} &
\colhead{$M_V${\tablenotemark{c}}} & 
\colhead{$V$} &
\colhead{$V_{\rm err}$} &
\colhead{$B-V$} &
\colhead{$(B-V)_{\rm err}$} &
\colhead{$V-I$} &
\colhead{$(V-I)_{\rm err}$} &
\colhead{Notes{\tablenotemark{d}}}
}
\startdata
37 & $  -10.63 $ &  24.06 &  0.06 &   0.16 &       0.07 &  0.52 &       0.09 & B \\
38 &   -----   &    23.57 &  0.04 &   1.30 &       0.08 &  2.78 &       0.04 & S \\
39 & $  -10.32 $ &  24.37 &  0.07 &   0.75 &       0.12 &  1.11 &       0.09 & C \\
40 & $  -10.29 $ &  24.40 &  0.07 &   0.24 &       0.09 &  0.46 &       0.11 & B \\
41 &   -----   &    23.04 &  0.03 &   1.19 &       0.05 &  2.12 &       0.03 & S \\
42 & $  -13.58 $ &  21.11 &  0.01 &   0.80 &       0.01 &  1.12 &       0.01 & D \\
43 & $   -9.69 $ &  25.00 &  0.13 &   0.13 &       0.16 &  0.72 &       0.20 & B \\
44 &   -----   &    23.33 &  0.03 &   1.25 &       0.07 &  1.83 &       0.04 & S \\
45 & $  -10.83 $ &  23.86 &  0.05 &   0.03 &       0.06 &  0.51 &       0.08 & A \\
46 & $   -9.96 $ &  24.73 &  0.09 &   0.20 &       0.11 &  0.33 &       0.17 & B \\
47 & $  -10.42 $ &  24.27 &  0.07 &   0.61 &       0.10 &  0.92 &       0.10 & C \\
48 & $  -10.38 $ &  24.31 &  0.08 &   0.28 &       0.11 &  0.11 &       0.16 & B \\
49 & $  -10.03 $ &  24.66 &  0.12 &   0.11 &       0.14 &  0.58 &       0.17 & B \\
50 & $  -12.90 $ &  21.79 &  0.01 &   1.01 &       0.02 &  1.32 &       0.02 & D \\
51 & $  -10.97 $ &  23.72 &  0.04 &   0.21 &       0.06 &  0.58 &       0.07 & B \\
52 & $  -11.57 $ &  23.12 &  0.03 &   0.50 &       0.04 &  0.73 &       0.04 & C \\
53 & $  -11.45 $ &  23.24 &  0.03 &   0.49 &       0.04 &  0.80 &       0.04 & C \\
54 & $  -13.50 $ &  21.19 &  0.01 &   0.29 &       0.02 &  0.88 &       0.02 & B \\
55 & $  -11.08 $ &  23.61 &  0.04 &   0.09 &       0.05 &  0.50 &       0.07 & A \\
56 & $  -10.40 $ &  24.29 &  0.07 &   0.89 &       0.11 &  0.90 &       0.09 & D \\
57 &   -----   &    21.90 &  0.01 &   1.34 &       0.02 &  2.12 &       0.01 & S \\
58 & $  -10.69 $ &  24.00 &  0.07 & $-0.12$&       0.08 &  0.39 &       0.14 & A \\
59 &   -----   &    23.54 &  0.06 &   1.14 &       0.11 &  1.84 &       0.06 & S \\
60 & $  -10.82 $ &  23.87 &  0.04 &   0.16 &       0.06 &  0.42 &       0.08 & B \\
61 & $  -10.36 $ &  24.33 &  0.06 &   0.93 &       0.11 &  1.16 &       0.08 & D \\
62 &   -----   &    24.54 &  0.08 &   1.26 &       0.18 &  1.77 &       0.09 & S \\
63 & $   -9.40 $ &  25.29 &  0.14 &   0.40 &       0.19 &  0.73 &       0.20 & C \\
64 & $  -13.23 $ &  21.46 &  0.01 &   0.84 &       0.02 &  1.16 &       0.01 & D \\
65 & $  -10.86 $ &  23.83 &  0.05 &   0.23 &       0.06 &  0.81 &       0.07 & B \\
66 & $  -13.04 $ &  21.64 &  0.01 &   0.89 &       0.02 &  1.19 &       0.01 & D \\
67 &   -----     &  23.50 &  0.04 &   1.36 &       0.07 &  1.40 &       0.04 & S \\
68 &   -----     &  24.81 &  0.10 &   1.12 &       0.18 &  2.71 &       0.10 & S \\
69 & $   -9.94 $ &  24.75 &  0.10 &   0.07 &       0.12 &  0.36 &       0.16 & A \\
\enddata
\tablenotetext{a}{All photometry has been corrected for Galactic extinction as described in 
$\S$~\ref{sec:finalphot}.}
\tablenotetext{b}{The numeric labels identify the sources as labeled in
Figure~\ref{fig:7319-panel}a.}
\tablenotetext{c}{ $M_V$ was calculated assuming a redshift distance modulus of 
34.69 corresponding to 
85~Mpc for $H_0=75$~km~s$^{-1}$~Mpc$^{-1}$ and $q_0=0.1$.  Note that the
absolute magnitudes of putative stars are marked with -----.}
\tablenotetext{d}{Letters correspond to $B-V$ color groups as defined in the
caption to Figure~{\ref{fig:tail-figs}}.}
\end{deluxetable}
%
%\clearpage
%--------------------------------------------------------------------------------------
%Table of photometric data for NGC 7318 and the Northern Star Formation Region.
%\normalsize
\begin{deluxetable}{rrccrcccl}
\tablewidth{0pt}
\tablecaption{Photometric Data: Northern Starburst Region and NGC~7318.{\tablenotemark{a}}
\label{tab:7318-table}}
\tablehead{
\colhead{Label{\tablenotemark{b}}} &
\colhead{$M_V${\tablenotemark{c}}} & 
\colhead{$V$} &
\colhead{$V_{\rm err}$} &
\colhead{$B-V$} &
\colhead{$(B-V)_{\rm err}$} &
\colhead{$V-I$} &
\colhead{$(V-I)_{\rm err}$} &
\colhead{Notes{\tablenotemark{d}}}
}
\startdata
 70 & $ -10.09 $ & 24.60 & 0.10 &  0.17 & 0.12 &  0.68 & 0.14 & B\\
 71 & $ -13.80 $ & 20.89 & 0.01 &  1.04 & 0.01 &  1.22 & 0.01 & D\\
 72 & $ -10.04 $ & 24.65 & 0.12 &$-0.10$& 0.14 &  0.75 & 0.17 & A\\
 73 & $ -10.66 $ & 24.03 & 0.07 &$-0.07$& 0.08 &  0.59 & 0.10 & A\\
 74 & $ -10.67 $ & 24.02 & 0.05 &  0.29 & 0.07 &  0.62 & 0.08 & B\\
%\\
 75 & $  -9.81 $ & 24.88 & 0.12 &  0.42 & 0.18 &  0.57 & 0.19 & C\\
 76 & $ -12.71 $ & 21.98 & 0.01 &  0.06 & 0.02 &  0.49 & 0.02 & A*\\
 77 & $ -10.38 $ & 24.31 & 0.07 &  0.70 & 0.11 &  0.76 & 0.10 & C\\
 78 & $ -12.41 $ & 22.28 & 0.02 &  0.07 & 0.03 &  0.47 & 0.03 & A*\\
 79 & $ -11.48 $ & 23.21 & 0.04 &$-0.01$& 0.05 &  0.49 & 0.06 & A*\\
%\\
 80 & $  -9.65 $ & 25.04 & 0.10 &  0.88 & 0.18 &  1.02 & 0.15 & D\\
 81 & $ -10.74 $ & 23.95 & 0.05 &  0.04 & 0.06 &  0.44 & 0.09 & A*\\
 82 & $ -10.61 $ & 24.08 & 0.06 &  0.20 & 0.07 &  0.57 & 0.10 & B\\
 83 & $ -10.10 $ & 24.59 & 0.12 &  0.22 & 0.14 &  0.32 & 0.19 & B\\
 84 &   -----  &   22.73 & 0.02 &  1.30 & 0.04 &  2.05 & 0.02 & S \\
%\\
 85 & $  -9.57 $ & 25.12 & 0.12 &$-0.19$& 0.15 &  0.62 & 0.19 & A\\
 86 & $ -10.01 $ & 24.68 & 0.11 &$-0.12$& 0.12 &  0.19 & 0.19 & A\\
 87 & $ -10.60 $ & 24.09 & 0.06 &  0.15 & 0.07 &  0.66 & 0.09 & B\\
 88 & $ -10.48 $ & 24.21 & 0.07 &  0.93 & 0.12 &  1.00 & 0.10 & D\\
 89 & $ -11.82 $ & 22.87 & 0.02 &$-0.10$& 0.03 &  0.20 & 0.05 & A\\
%\\
 90 & $ -13.01 $ & 21.68 & 0.01 &$-0.16$& 0.02 &  0.06 & 0.03 & A\\
 91 & $ -12.14 $ & 22.55 & 0.02 &  0.03 & 0.03 &  0.42 & 0.03 & A*\\
 92 & $ -10.75 $ & 23.94 & 0.05 &  0.29 & 0.07 &  0.38 & 0.09 & B\\
 93 & $ -13.43 $ & 21.26 & 0.01 &$-0.13$& 0.01 &$-0.02$& 0.02 & A\\
 94 & $ -10.33 $ & 24.36 & 0.08 &$-0.07$& 0.09 &  0.65 & 0.12 & A\\
%\\
 95 & $ -11.12 $ & 23.57 & 0.05 &$-0.23$& 0.05 &  0.04 & 0.09 & A\\
 96 & $ -10.86 $ & 23.83 & 0.10 &  0.74 & 0.15 &  1.01 & 0.15 & C\\
 97 & $ -10.39 $ & 24.30 & 0.07 &  0.31 & 0.09 &  0.51 & 0.11 & C\\
 98 & $ -10.34 $ & 24.35 & 0.07 &  0.30 & 0.09 &  0.47 & 0.11 & C\\
 99 & $  -9.86 $ & 24.83 & 0.15 &  0.25 & 0.19 &  1.18 & 0.20 & B\\
%\\
100 & $ -11.26 $ & 23.43 & 0.04 &  0.26 & 0.05 &  0.44 & 0.06 & B\\
101 & $ -12.58 $ & 22.11 & 0.02 &  0.36 & 0.03 &  0.86 & 0.03 & C, DG\\
102 & $ -10.32 $ & 24.37 & 0.07 &  0.04 & 0.09 &  0.41 & 0.12 & A*\\
103 & $ -10.31 $ & 24.38 & 0.09 &  0.61 & 0.14 &  1.19 & 0.13 & C\\
104 & $ -13.52 $ & 21.17 & 0.01 &  0.41 & 0.02 &  0.61 & 0.02 & C, DG\\
%\\
105 &   -----    & 22.74 & 0.08 &  1.34 & 0.16 &  1.39 & 0.10 & S\\
106 & $ -10.53 $ & 24.16 & 0.10 &  0.86 & 0.16 &  1.34 & 0.12 & D, DG\\
107 & $ -10.60 $ & 24.09 & 0.08 &  0.82 & 0.13 &  1.02 & 0.11 & D\\
108 & $ -11.21 $ & 23.48 & 0.08 &  0.92 & 0.15 &  1.30 & 0.10 & D, DG\\
109 & $ -12.50 $ & 22.19 & 0.02 &  0.40 & 0.03 &  0.82 & 0.03 & C, DG\\
%\\
110 & $ -11.60 $ & 23.09 & 0.04 &  0.47 & 0.06 &  0.84 & 0.06 & C, DG\\
111 & $ -10.05 $ & 24.64 & 0.09 &  0.45 & 0.12 &  1.08 & 0.12 & C\\
112 & $ -10.23 $ & 24.46 & 0.07 &  0.43 & 0.10 &  1.39 & 0.09 & C\\
113 & $ -10.85 $ & 23.84 & 0.04 &$-0.35$& 0.05 &  0.22 & 0.08 & A\\
114 & $ -10.67 $ & 24.02 & 0.07 &  0.04 & 0.09 &  0.36 & 0.12 & A*, DG\\
%\\
115 &   -----  &   23.00 & 0.03 &  1.13 & 0.05 &  2.26 & 0.03 & S \\
116 &   -----  &   23.68 & 0.05 &  1.64 & 0.13 &  2.25 & 0.05 & S \\
117 &   -----  &   20.90 & 0.01 &  1.29 & 0.01 &  1.75 & 0.01 & S \\
118 & $ -11.06 $ & 23.63 & 0.05 &$-0.23$& 0.05 &  0.09 & 0.09 & A, DG\\
119 &   -----  &   23.34 & 0.03 &  1.19 & 0.07 &  1.30 & 0.04 & S\\
%\\
120 & $ -10.67 $ & 24.02 & 0.07 &  0.04 & 0.08 &  0.40 & 0.10 & A*\\
121 & $ -11.87 $ & 22.82 & 0.02 &$-0.10$& 0.03 &  0.16 & 0.04 & A\\
122 &   -----  &   23.88 & 0.05 &  1.25 & 0.11 &  2.20 & 0.05 & S \\
123 & $ -11.37 $ & 23.32 & 0.05 &$-0.18$& 0.07 &$-0.02$& 0.10 & A\\
124 & $ -10.25 $ & 24.44 & 0.08 &  0.56 & 0.12 &  1.09 & 0.12 & C\\
%\\
125 & $ -10.08 $ & 24.61 & 0.10 &  0.57 & 0.14 &  1.12 & 0.12 & C\\
126 & $ -13.34 $ & 21.34 & 0.01 &  0.75 & 0.01 &  1.02 & 0.01 & C\\
127 &   -----  &   22.25 & 0.02 &  1.39 & 0.03 &  1.72 & 0.02 & S \\
128 & $ -10.52 $ & 24.17 & 0.10 &  0.85 & 0.15 &  1.49 & 0.13 & D\\
129 &   -----  &   22.37 & 0.02 &  1.31 & 0.04 &  2.15 & 0.02 & S \\
%\\
130 & $ -10.62 $ & 24.07 & 0.08 &$-0.11$& 0.10 &  0.56 & 0.13 & A\\
131 & $ -11.56 $ & 23.13 & 0.05 &$-0.20$& 0.06 &  0.21 & 0.09 & A\\
132 &   -----  &   22.28 & 0.02 &  1.29 & 0.04 &  2.35 & 0.02 & S \\
133 & $  -9.65 $ & 25.04 & 0.14 &  0.59 & 0.20 &  0.97 & 0.19 & B\\
134 & $ -12.83 $ & 21.86 & 0.01 &$-0.01$& 0.02 &  0.52 & 0.02 & A*\\
%\\
135 & $ -11.58 $ & 23.11 & 0.04 &$-0.19$& 0.04 &$-0.15$& 0.09 & A\\
136 & $ -10.36 $ & 24.33 & 0.10 &  0.30 & 0.15 &  0.31 & 0.17 & C\\
137 & $ -12.43 $ & 22.26 & 0.04 &  0.00 & 0.05 &  0.44 & 0.05 & A*\\
138 & $ -10.48 $ & 24.21 & 0.09 &  0.64 & 0.13 &  0.67 & 0.16 & C\\
139 & $  -9.80 $ & 24.89 & 0.12 &  0.63 & 0.16 &  1.10 & 0.15 & C\\
\enddata
\tablenotetext{a}{All photometry has been corrected for Galactic extinction as described in 
$\S$~\ref{sec:finalphot}.}
\tablenotetext{b}{The numeric labels identify the sources as labeled in
Figure~\ref{fig:7318-panel}a.}
\tablenotetext{c}{ $M_V$ was calculated assuming a redshift distance modulus of 
34.69 corresponding to 
85~Mpc for $H_0=75$~km~s$^{-1}$~Mpc$^{-1}$ and $q_0=0.1$.  Note that the
absolute magnitudes of putative stars are marked with -----.}
\tablenotetext{d}{Letters correspond to $B-V$ color groups as defined in the
caption to Figure~{\ref{fig:tail-figs}}.  An asterisk (*) indicates a source
in the tight color group with ages consistent with $~7$~Myr (see $\S$~3.4 for discussion).  
The star cluster candidates in the 
dwarf galaxy in the northern starburst region (Figure~{\ref{fig:dwarf-fig}})
are designated with the letters DG.}
\end{deluxetable}
\clearpage


\begin{thebibliography}{XXX}

\bibitem[Ashman \& Zepf(1992)]{ashman92}
Ashman, K. M. \& Zepf, S. E. 1992, \apj, 384, 50

\bibitem[Bruzual \& Charlot(1993)]{bruzual}
Bruzual, G. A. \& Charlot, S. 1993, \apj, 405, 538

\bibitem[Elmegreen \etal(2000)]{elmegreen2000} 
Elmegreen, B.\ G., Efremov, Y., 
Pudritz, R.\ E., \& Zinnecker, H.\ 2000, Protostars and Planets IV (Book - 
Tucson: University of Arizona Press; eds Mannings, V., Boss, A.P., Russell, 
S.\ S.), p.\ 179 

\bibitem[Gnedin \& Ostriker(1997)]{gnedin97}
Gnedin, O. Y. \& Ostriker, J. P 1997, \apj, 474, 223

\bibitem[Harris(1991)]{harris91}
Harris, W. E. 1991, \araa, 29, 543

\bibitem[Hibbard \& Mihos(1995)]{hibbard95}
Hibbard, J. \& Mihos, C. 1991, \aj, 110, 140

\bibitem[Hickson(1982)]{hickson82}
Hickson, P. 1982, \apj, 255, 382

\bibitem[Hickson \& Mendes de Oliveira(1992)]{hickson92}
Hickson, P. \& Mendes de Oliveira, C. 1992, \apj, 399, 353

\bibitem[Hill \etal(1998a)]{lrs1}
Hill, G. J., Nicklas, H. E., MacQueen, P. J., Tejada, C., Cobos Duenas, F. J.,
\& Mitsch, W. 1998a, Proc. SPIE, 3355, 375

\bibitem[Hill \etal(1998b)]{lrs2}
Hill, G. J., Nicklas, H. E., MacQueen, P. J., Mitsch, W., Wellem, W., Altman, W., Wesley, G. L.,
\& Ray, F. B.  1998b, Proc. SPIE, 3355, 433

\bibitem[Holtzman \etal(1992)]{Holtz92}
Holtzman \etal 1992, AJ, 102, 691

\bibitem[Holtzman \etal(1995)]{Holtz95}
Holtzman, J. A., Burrows, C. J., Casertano, S., Hester, J. J., 
Trauger, J. T., Watson, A. M., \& Worthey, G. 1995, \pasp, 107, 1065

\bibitem[Hunsberger, Charlton, \& Zaritsky(1996)]{sally96}
Hunsberger, S. D., Charlton, J. C. \& Zaritsky, D.  1996, \apj, 462, 50 

\bibitem[Hunsberger, Charlton, \& Zaritsky(1998)]{sally98}
Hunsberger, S. D., Charlton, J. C. \& Zaritsky, D.  1998 \apj, 505, 536

\bibitem[Johnson \& Conti(2000)]{johnson00}
Johnson, K. E., \& Conti, P. S. 2000, \aj, 119, 2146

\bibitem[Johnson \etal(1999)]{johnson99}
Johnson, K. E., Vacca, W. D., Leitherer, C., Conti, P. S., \& Lipscy, S. J. 1999,
\aj, 117, 1708

\bibitem[Knierman \etal(2001)]{knierman01}
Knierman, K. A., Gallagher, S. C., Charlton, J. C., Hunsberger, S. D., Whitmore, B.,
Kundu, A., Hibbard, J., \& Zaritsky, D. 2001, \aj, submitted

\bibitem[Kundu \etal(1999)]{kundu99}
Kundu, A., Whitmore, B. C., Sparks, W. B., \& Macchetto, F. D., Zepf, S. E.,
\& Ashman, K. M. 1999, \aj, 513, 733

\bibitem[Mendes de Oliveira \& Hickson(1994)]{claudia94}
Mendes de Oliveira, C., \& Hickson, P. 1994, \apj, 427, 684

\bibitem[Mirabel, Dottori, \& Lutz(1992)]{mirabel}
Mirabel, I. F., Dottori. H., \& Lutz, D. 1992, \aap, 256, L19

\bibitem[Miller \etal(1997)]{miller97}
Miller, B. W., Whitmore, B. C., Schweizer, F. \& Fall, S. M. 1997, \aj, 114, 2381

\bibitem[Moles, Sulentic, \& M\'arquez(1997)]{moles97}
Moles, M., Sulentic, J. W., \& M\'arquez, I. 1997, \apj, 485, L69 (MSM97)

\bibitem[Moles, M\'arquez, \& Sulentic(1998)]{moles98}
Moles, M., M\'arquez, I., \& Sulentic, J. W. 1998, \aap, 334, 473

\bibitem[Paturel \etal(1997)]{paturel}
Paturel, G., et al. 1997, \aaps, 124, 109 

\bibitem[Pietsch \etal(1997)]{pietsch}
Pietsch, W., Trinchieri, G., Arp, H., \& Sulentic, J. W. 1997, \aap, 322, 89

\bibitem[Ponman \etal(1996)]{ponman96}
Ponman, T. J., Bourner, P. D. J., Ebeling, H., \& Bohringer, H. 1996,
\mnras, 283, 690

\bibitem[Ramsey \etal(1998)]{het}
Ramsey, L. W., \etal 1998, Proc. SPIE, 3352, 34

\bibitem[Reed, Hesser, \& Shawl(1988)]{reed88}
Reed, B. C., Hesser, J. E., \& Shawl, S. J. 1988, \pasp, 100, 545

\bibitem[Saracco \& Ciliegi(1995)]{saracco95}
Saracco, P. \& Ciliegi, P. 1995, \aap, 301, 348

\bibitem[Schombert, Wallin, \& Struck-Marcell(1990)]{schombert90}
Schombert, J. M., Wallin, J. F. \& Struck-Marcell, C.  1990, \aj, 99, 497 

\bibitem[Seaton(1979)]{seaton79}
Seaton, M.J. 1979, \mnras, 187, 73

\bibitem[Schweizer \etal(1996)]{schweizer96}
Schweizer, F., Miller, B., Whitmore, B., \& Fall, S.M. 1996, AJ, 112, 1839

\bibitem[Shostak, Allen, \& Sullivan(1984)]{shostak}
Shostak, G. S., Allen, R. J., \& Sullivan, W. T.(1984, \aap, 139, 15

\bibitem[Smith \& Struck(2000)]{smith}
Smith, B. J., \& Struck, C. 2000, \aj, in press

\bibitem[Stephan(1877)]{stephan}
Stephan, M. E. 1877, CR Acad. Sci. Paris, 84, 641

\bibitem[Stetson(1987)]{stetson}
Stetson, P.B. 1987, \pasp, 99, 191

\bibitem[van der Hulst \& Rots (1981)]{vdhulst}
van der Hulst, J. M. \& Rots, A. H. 1981, \aj, 86, 12

\bibitem[Verdes-Montenegro \etal(1998)]{verdes98}
Verdes-Montenegro, L., Yun, M. S., Perea, J.,
del Olmo, A., \& and Ho, P. T. P. 1998, \apj, 497, 89

\bibitem[V{\'{\i}}lchez \& Iglesias-P\'aramo(1998)]{vilchez98}
V{\'{\i}}lchez, J. M. \& Iglesias-P\'aramo, J. 1998, \apjs, 117, 1

\bibitem[Whitmore, Heyer \& Casertano(1999)]{whc99} 
Whitmore, B., Heyer, I., \& Casertano, S. 1999, \pasp, 111, 1559 

\bibitem[Whitmore \etal(1993)] {whitmore93}
Whitmore, B., Schweizer, F., Leitherer, C., Borne, K., \& Robert,
C. 1993, \aj, 106, 1354

\bibitem[Whitmore \etal(1999)] {whitmore99}
Whitmore, B., Zhang, Q., Leitherer, C., Fall, S.M., Schweizer, F., \&
Miller, B. 1999, \aj, 118, 1551

\bibitem[Xu, Sulentic, \& Tuffs(1999)]{xu99}
Xu, C. , Sulentic, J. W. \& Tuffs, R.  1999, \apj, 512, 178

\bibitem[Zepf \& Ashman(1993)]{zepf93}
Zepf, S. E., \& Ashman, K. M. 1993, \mnras, 264, 611

\bibitem[Zepf \etal(1999)] {zepf99}
Zepf, S., Ashman, K., English, J., Freeman, K., \& Sharples, R. 1999, AJ, 118, 752

\bibitem[Zhang \& Fall(1999)]{zhang99}
Zhang, Q. \& Fall, S. M. 1999, ApJ, 527, L81


\end{thebibliography}
\end{document}